\documentclass[twocolumn,aps,pra,amsmath,amssymb]{revtex4-1}
\usepackage[latin9]{inputenc}
\usepackage{amsmath}
\usepackage{amssymb}
\usepackage{graphicx}
\usepackage{wasysym}

\makeatletter
\usepackage{amssymb}
\usepackage{epsfig}
\usepackage{bm}
\usepackage{dcolumn}
\usepackage{color}

\makeatother

\begin{document}

\title{Low-momentum dynamic structure factor of a strongly interacting Fermi
gas at finite temperature: A two-fluid hydrodynamic description}

\author{Hui Hu$^{1}$, Peng Zou$^{2}$, and Xia-Ji Liu$^{1}$}

\affiliation{$^{1}$Centre for Quantum and Optical Science, Swinburne University
of Technology, Melbourne, Victoria 3122, Australia}

\affiliation{$^{2}$College of Physics, Qingdao University, Qingdao 266071, China}

\date{\today}
\begin{abstract}
We provide a description of the dynamic structure factor of a homogeneous
unitary Fermi gas at low momentum and low frequency, based on the
dissipative two-fluid hydrodynamic theory. The viscous relaxation
time is estimated and is used to determine the regime where the hydrodynamic
theory is applicable and to understand the nature of sound waves in
the density response near the superfluid phase transition. By collecting
the best knowledge on the shear viscosity and thermal conductivity
known so far, we calculate the various diffusion coefficients and
obtain the damping width of the (first and second) sounds. We find
that the damping width of the first sound is greatly enhanced across
the superfluid transition and very close to the transition the second
sound might be resolved in the density response for the transferred
momentum up to the half of Fermi momentum. Our work is motivated by
the recent measurement of the local dynamic structure factor at low
momentum at Swinburne University of Technology and the on-going experiment
on sound attenuation of a homogeneous unitary Fermi gas at Massachusetts
Institute of Technology. We discuss how the measurement of the velocity
and damping width of the sound modes in low-momentum dynamic structure
factor may lead to an improved determination of the universal superfluid
density, shear viscosity and thermal conductivity of a unitary Fermi
gas.
\end{abstract}

\pacs{03.75.Ss, 67.85.Lm, 05.30.Fk, 67.85.De}
\maketitle

\section{Introduction}

Dynamic structure factor, which determines the response of the system
with respect to an external density perturbation, provides valuable
information on the low-energy elementary excitations of strongly correlated
many-body systems \cite{Nozieres1990,Griffin1993}. In superfluid
$^{4}$He, it was measured by Brillouin and Raman light scattering
(at long wave-length) \cite{Woods1973} and inelastic neutron scattering
\cite{Griffin1993} and played a central role in consolidating the
modern ideas of quasiparticles (i.e., phonons, maxons and rotons)
\cite{Khalatnikov1965} and of two-fluid hydrodynamics of superfluids
\cite{Tisza1938,Landau1941,Khalatnikov1965}.

A strongly interacting Fermi gas of ultracold atoms with infinitely
large scattering length (i.e., $a_{s}=\pm\infty$, the so-called unitary
limit) is a new type of Fermi superfluids \cite{Bloch2008}, which
has received increasing attention from different branches of physics
since its first realization in 2002 \cite{OHara2002}. To date, a
number of intriguing properties of a unitary Fermi gas \cite{Randeria2014},
in particular the universal equations of state \cite{Ku2012}, have
been measured to a great accuracy. However, less is known about its
low-energy elementary excitation spectrum \cite{Joseph2007,Sidorenkov2013}.
This is largely due to the external harmonic trapping potential that
is necessary to keep atoms from escaping. Only very recently, two-photon
Bragg spectroscopy has been applied with low transferred momentum
to determine the \emph{local} dynamic structure factor of the central
part of a trapped unitary Fermi gas at low energy \cite{Hoinka2017},
and hence visualize the Bogoliubov-Anderson phonon mode in the deep
superfluid phase at $T=0.09(1)T_{F}\simeq0.54T_{c}$, where $T_{F}$
is the Fermi temperature and $T_{c}\simeq0.167T_{F}$ is the superfluid
transition temperature \cite{Ku2012}. Analogous to inelastic neutron
scattering in superfluid $^{4}$He \cite{Griffin1993}, it is natural
to anticipate that further measurements of the temperature dependence
of the local low-momentum Bragg spectroscopy may give a full plot
of the dynamic structure factor and elementary excitation spectrum
across the superfluid transition. At this point, it is worth noting
that the measurement of sound propagation of a unitary Fermi gas trapped
in a \emph{uniform} box potential is also under-going \cite{Zwierlein2017}.
Analogous to sound attenuation experiments in superfluid $^{4}$He
\cite{Griffin1993}, the determination of sound velocity and attenuation
of a uniform unitary Fermi gas provides a useful alternative way to
characterize its elementary excitations.

These on-going experiments on density response at Swinburne University
Technology (SUT) and at Massachusetts Institute of Technology (MIT)
urge us to look for new theoretical development to describe the dynamic
structure factor of a unitary Fermi gas at low momentum and low energy.
This turns out to be a notoriously difficult task, since for a unitary
Fermi gas there is no small parameter to control the accuracy of the
theory \cite{Randeria2014}. Fortunately, under certain conditions
at low energy the density response of a unitary Fermi gas may be well
described by the seminal Landau two-fluid hydrodynamic theory \cite{Khalatnikov1965}.
The requirement for hydrodynamics is usually summarized as $\omega\tau\ll1$,
where $\omega$ is the frequency of a collective excitation and $\tau$
is the appropriate relaxation time. The short relaxation time or collision
time for excitations ensures the establishment of local hydrodynamic
equilibrium and the dynamics of the system is thus governed by a set
of equations that satisfy conservation laws \cite{Khalatnikov1965}.
In the previous theoretical investigations \cite{Arahata2009,Hu2010},
the non-dissipative two-fluid hydrodynamic theory has been applied
to illustrate the sizable coupling between first and second sound
in a unitary Fermi gas and to show the promising opportunity of exciting
second sound with density probes.

The purpose of the present work is to update such a two-fluid hydrodynamic
calculation by including the crucial \emph{dissipation} terms. In
doing so, we are able to obtain quantitative predictions of the dynamic
structure factor in the hydrodynamic regime, once the relevant transport
coefficients in the dissipation terms are known. In greater detail,
we estimate the realistic experimental condition for reaching the
hydrodynamic regime and clarify the nature of sound waves of a unitary
Fermi gas near the superfluid transition. We then calculate the hydrodynamic
dynamic structure factor based on the existing knowledge on transport
coefficients such as shear viscosity and thermal conductivity. We
also discuss the interesting features of the dynamic structure factor
that may arise in the on-going experiments at SUT, with emphasis on
the optimal experimental condition for probing the second sound.

It should be emphasized that the attempt to calculate the dynamic
structure factor with the use of the dissipative two-fluid hydrodynamic
theory, as the one carried out in the present work, can only be made
possible very recently, since the key inputs to the theory - the transport
coefficients of the unitary Fermi gas such as the shear viscosity
- are determined with reasonable precision only recently \cite{Joseph2015,Bluhm2017}.
Interestingly, we may reverse this procedure and consider the accurate
measurement of the dynamic structure factor as the input. The transport
coefficients can then be extracted from the measured velocity and
damping width of sounds. For this purpose, in the end of this paper,
we will discuss the sensitivity of the velocity and damping width
of sounds on the transport coefficients.

It is also worth noting that the sound attenuation of a strongly interacting
Fermi gas - which is more relevant to the on-going sound attenuation
experiment at MIT - has been theoretically investigated by Braby,
Chao and Schäfer \cite{Braby2010}, by taking a high-temperature approximation
for both shear viscosity and thermal conductivity (see Fig. 6 in Ref.
\cite{Braby2010}). In this work, we work with more accurate shear
viscosity (as recently measured) and focus on the more fundamental
property of dynamic structure factor.

The rest of the paper is set out as follows. In the next section,
we brief review the well-known dissipative two-fluid hydrodynamic
theory and present the expression of the dynamic structure factor.
The various input parameters to the theory are discussed. In Sec.
III, we estimate the viscous relaxation time and determine the conditions
for the application of the hydrodynamic theory. The nature of sound
waves at different temperatures near the superfluid transition is
sketched. In Sec. IV, we report in detail the hydrodynamic dynamic
structure factor at two typical transferred momenta, considering the
realistic experimental situations at SUT and MIT, respectively. We
focus particularly on the dependence of the second sound on the transferred
momentum. In Sec. V, we discuss the dependence of the sound velocity
and sound attenuation on the input parameters of superfluid density
and thermal conductivity. Finally, we conclude in Sec. VI. 

\section{Two-fluid hydrodynamic theory}

The dynamic structure factor $S(\mathbf{k},\omega)$ measures the
scattering rate of a density probe that imports a momentum $\hbar\mathbf{k}$
and energy $\hbar\omega$ to the system \cite{Nozieres1990,Griffin1993}.
It is formally related to the imaginary part of the density response
function $\chi_{nn}\left(\mathbf{k},\omega+i0^{+}\right)$ by
\begin{equation}
S\left(\mathbf{k},\omega\right)=-\frac{1}{n\pi}\frac{1}{\left(1-e^{-\beta\hbar\omega}\right)}\textrm{Im}\chi_{nn}\left(\mathbf{k},\omega+i0^{+}\right),\label{eq:DynamicStructureFactor}
\end{equation}
where $n$ is the total number density and $\beta=1/(k_{B}T)$ is
the inverse temperature. The expression of the density response function
within the two-fluid hydrodynamic theory was first derived by Hohenberg
and Martin in their seminal work \cite{Hohenberg1965} and takes the
following form for an isotropic superfluid (such that $\chi_{nn}(\mathbf{k},\omega)=\chi_{nn}(k,\omega)$),
\begin{equation}
\chi_{nn}=\frac{\left(nk^{2}/m\right)\left(\omega^{2}-v^{2}k^{2}+iD_{s}k^{2}\omega\right)}{\left(\omega^{2}-c_{1}^{2}k^{2}+iD_{1}k^{2}\omega\right)\left(\omega^{2}-c_{2}^{2}k^{2}+iD_{2}k^{2}\omega\right)}.\label{eq:DensityReponseFunction}
\end{equation}
Here, $m$ is the mass of atoms, 
\begin{equation}
v^{2}\equiv T\frac{s^{2}}{c_{v}}\frac{\rho_{s}}{\rho_{n}}
\end{equation}
is a velocity determined by the equilibrium entropy per unit mass
$s\equiv S/(Nm)$, the specific heat per unit mass $c_{v}\equiv T(\partial s/\partial T)_{\rho}$
and the superfluid and normal fluid mass densities, $\rho_{s}$ and
$\rho_{n}$ (the total mass density is $\rho=mn=\rho_{s}+\rho_{n}$).
$c_{1}$ and $c_{2}$ are the well-known exact first and second sound
velocities that satisfy the relations,
\begin{eqnarray}
c_{1}^{2}+c_{2}^{2} & = & v^{2}+v_{s}^{2},\label{eq: SoundVelocity1}\\
c_{1}^{2}c_{2}^{2} & = & v_{T}^{2}v_{s}^{2}=\frac{v^{2}v_{s}^{2}}{\gamma},\label{eq: SoundVelocity2}
\end{eqnarray}
where $v_{s}^{2}\equiv(\partial P/\partial\rho)_{s}$ and $v_{T}^{2}=(\partial P/\partial\rho)_{T}$
are the adiabatic and isothermal sound velocities, respectively. According
to standard thermodynamic relations, the ratio of these two velocities
is related to the ratio of two specific heats,
\begin{equation}
\gamma\equiv\frac{c_{p}}{c_{v}}=\frac{v_{s}^{2}}{v_{T}^{2}}.
\end{equation}
Quite generally, $v_{s}$ differs from $v_{T}$ due to the finite
thermal expansion of the system, implying that $\gamma>1$. It will
become clear later that this difference, as measured by the so-called
Landau-Placzek (LP) ratio $\epsilon_{\textrm{LP}}=\gamma-1$, determines
the coupling strength between first and second sound. In Eq. (\ref{eq:DensityReponseFunction}),
the damping rate or sound attenuation of the density response is characterized
by three diffusion coefficients $D_{1}$, $D_{2}$ and $D_{s}$, which
are determined by solving \cite{Hohenberg1965},
\begin{eqnarray}
D_{1}+D_{2} & = & \frac{4}{3}\frac{\eta}{\rho}+\frac{\kappa}{\rho c_{v}},\label{eq: DC1}\\
\frac{c_{1}^{2}D_{2}+c_{2}^{2}D_{1}}{v_{s}^{2}} & = & \frac{4}{3}\frac{\eta}{\rho}\left[\frac{v^{2}}{v_{s}^{2}}-\frac{2v^{2}}{\rho s}\frac{\left(\frac{\partial P}{\partial T}\right)_{\rho}}{v_{s}^{2}}+\frac{\rho_{s}}{\rho_{n}}\right]+\frac{\kappa}{\rho c_{p}},\label{eq: DC2}\\
D_{s} & = & \frac{4}{3}\frac{\eta}{\rho}\frac{\rho_{s}}{\rho_{n}}+\frac{\kappa}{\rho c_{v}},\label{eq: DC3}
\end{eqnarray}
where $\kappa$ is the thermal conductivity and $\eta$ is the shear
viscosity. In the above expressions, we do not include the various
second viscosities $\zeta_{i}$ ($i=1$,$2$,$3$,$4$), since for
a unitary Fermi gas, it is known that only $\zeta_{3}$ can be nonzero
but its value is too small to have sizable contribution \cite{Son2007,Escobedo2009}.

Eq. (2) gives the density response for any weakly or strongly interacting
superfluids that satisfy the Galilean invariance in the hydrodynamic
regime. This universal description is very powerful, considering the
lack of microscopic theoretical treatments in the strongly correlated
regime. The only inputs to the theory are the knowledge on the equations
of state, superfluid density and some transport coefficients like
shear viscosity and thermal conductivity. For a strongly correlated
unitary Fermi gas, all of them can be expressed in terms of some universal
functions that depend on a single parameter $T/T_{F}$ only, as first
suggested by Ho in his seminal universality work \cite{Ho2004}. Luckily,
we now start to have reliable results of these universal functions,
due to the endless efforts from both experimental and theoretical
sides. In particular, precise data of the equations of state have
been obtained at MIT in 2012 within a few percent accuracy \cite{Ku2012}.
The superfluid density has been determined at Innsbruck in 2013 from
the velocity of second sound that propagates along a highly elongated
harmonic trap \cite{Sidorenkov2013}. The shear viscosity of the unitary
Fermi gas has also been measured at North Carolina State University
(NCSU) in 2015 from the anisotropic expansion of the Fermi cloud for
a wide temperature window, ranging from $\sim T_{F}$ down to far
below the transition temperature \cite{Joseph2015}. These rapid experimental
advances make it possible to predict the dynamic structure factor
in the hydrodynamic regime by using the dissipative two-fluid hydrodynamic
theory.

\subsection{The sound velocities and diffusion coefficients}

In this work, we use the MIT equations of state \cite{Ku2012} and
the NCSU shear viscosity \cite{Joseph2015,Bluhm2017}. For the superfluid
density, unless specified otherwise we adopt the prediction from a
gaussian pair fluctuation (GPF) theory \cite{Hu2006,Fukushima2007,Taylor2008},
which agrees well with the Innsbruck data (see the inset of Fig. \ref{fig10_soundvelocityns}).
The thermal conductivity of a unitary Fermi gas, on the other hand,
is less investigated both experimentally and theoretically. We take
the known expression of the thermal conductivity at high temperature
\cite{Braby2010}, 
\begin{equation}
\frac{\kappa}{n\hbar}=\frac{k_{B}}{m}\frac{675\sqrt{2}\pi^{3/2}}{512}\vartheta^{3/2},\label{eq: ThermalConductivityHighTemperature}
\end{equation}
where $\vartheta\equiv T/T_{F}$ is the reduced temperature, and as
the first-order approximation, we assume that this expression is applicable
at all temperatures.

In greater detail, we express the pressure of a unitary Fermi gas
in terms of the universal energy function $f_{E}(\vartheta)\equiv E/(NE_{F})$,
the entropy in terms of $f_{s}(\vartheta)\equiv S/(Nk_{B})$, and
the shear viscosity in terms of $f_{\eta}(\vartheta)\equiv\eta/(n\hbar)$.
These dimensionless universal functions can be directly read from
the experimental data. For the thermal conductivity, we instead use
the dimensionless Prandtl ratio \cite{Braby2010},
\begin{equation}
\textrm{Pr}\left(\vartheta\right)\equiv\frac{\eta c_{p}}{\kappa}.
\end{equation}
Using the universal relation $P=2E/(3V)$ \cite{Ho2004,Thomas2005},
it is straightforward to obtain,
\begin{eqnarray}
v_{s}^{2} & = & \left(\frac{\partial P}{\partial\rho}\right)_{s}=\frac{5}{9}f_{E}\left(\vartheta\right)v_{F}^{2},\label{eq: vs2}\\
v_{T}^{2} & = & \left(\frac{\partial P}{\partial\rho}\right)_{T}=\left[\frac{5}{9}f_{E}\left(\vartheta\right)-\frac{2}{9}\vartheta f_{E}'\left(\vartheta\right)\right]v_{F}^{2}.\label{eq: vT2}
\end{eqnarray}
where $f'(\vartheta)\equiv df(\vartheta)/d\vartheta$ and $v_{F}$
is the Fermi velocity. Therefore, $\gamma$ and the LP ratio $\epsilon_{\textrm{LP}}\equiv\gamma-1$
are given by, 
\begin{eqnarray}
\frac{1}{\gamma} & = & \frac{v_{s}^{2}}{v_{T}^{2}}=1-\frac{2}{5}\theta\frac{f_{E}'\left(\vartheta\right)}{f_{E}\left(\vartheta\right)},\\
\epsilon_{\textrm{LP}} & = & \frac{\theta f_{E}'\left(\vartheta\right)}{\left(5/2\right)f_{E}\left(\theta\right)-\theta f_{E}'\left(\vartheta\right)}.
\end{eqnarray}
Using $s=(k_{B}/m)f_{s}(\vartheta)$, the specific heat $c_{v}$ takes
the form, 
\begin{equation}
c_{v}=\frac{k_{B}}{m}\vartheta f_{s}'\left(\vartheta\right).
\end{equation}
We then obtain,
\begin{equation}
v^{2}=\frac{1}{2}\frac{n_{s}}{n_{n}}\frac{f_{s}^{2}\left(\vartheta\right)}{f_{s}'\left(\vartheta\right)}v_{F}^{2},\label{eq: v2}
\end{equation}
where $n_{s}\equiv\rho_{s}/\rho$ and $n_{n}\equiv\rho_{n}/\rho=1-n_{s}$
are the superfluid fraction and normal fluid fraction, respectively.

\begin{figure}
\centering{}\includegraphics[width=0.48\textwidth]{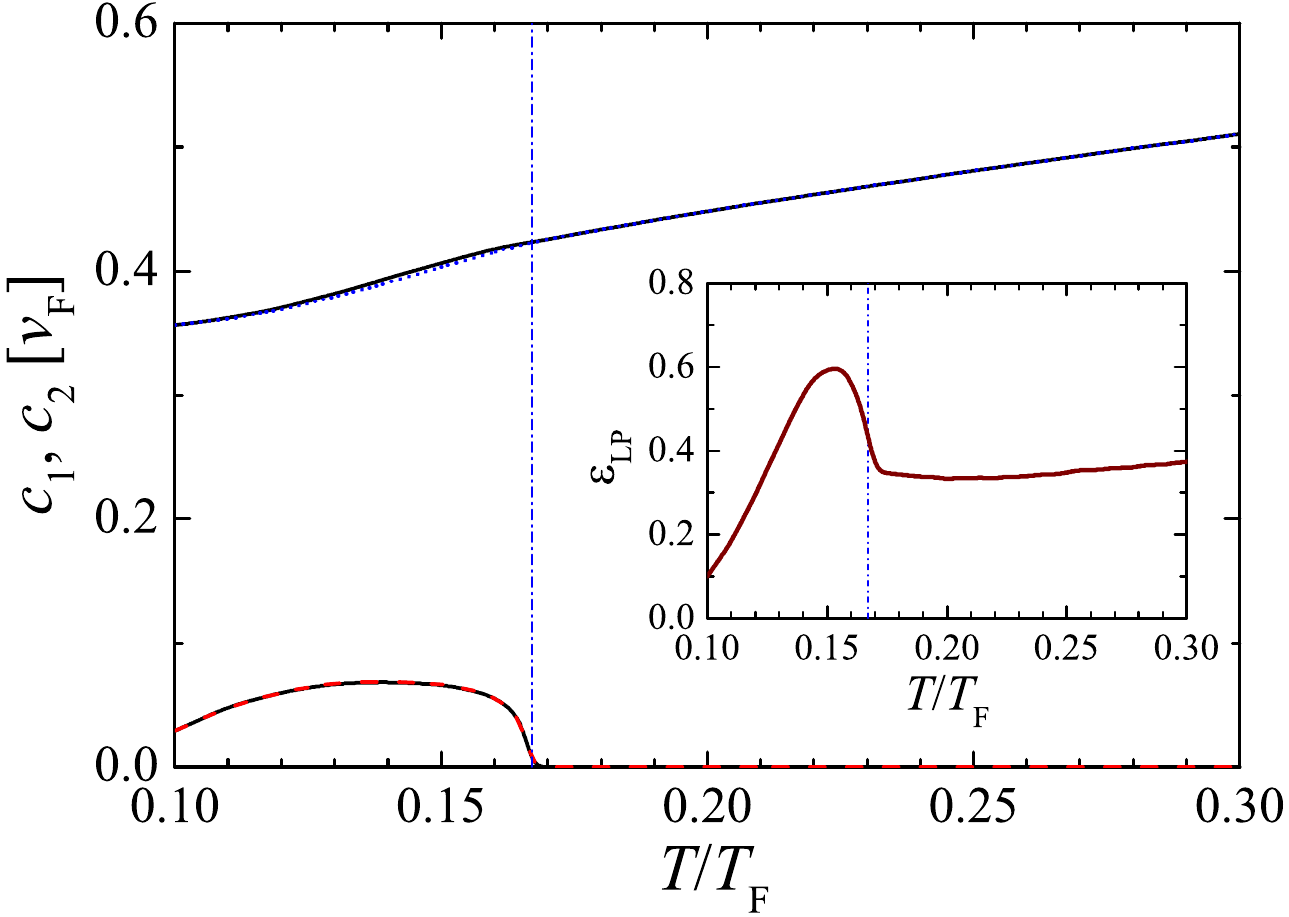}\caption{\label{fig1_soundvelocity} (color online). First and second sound
velocities of a unitary Fermi gas as a function of the reduced temperature
$\vartheta=T/T_{F}$. The blue dash-dotted line and the red dashed
line are the approximate results $c_{1}=v_{s}$ and $c_{2}=v/\sqrt{\gamma}$,
which are nearly indistinguishable with the exact results (black curves)
due to the small LP ratio. The latter is shown in the inset. Here,
we adopt the MIT equations of state \cite{Ku2012} and the GPF superfluid
density \cite{Taylor2008}. In this figure and some figures below,
the (vertical) thin blue dot-dashed line indicates the superfluid
transition temperature $T_{c}\simeq0.167T_{F}$. }
\end{figure}

By plugging the expressions (\ref{eq: vs2})-(\ref{eq: v2}) into
Eqs. (\ref{eq: SoundVelocity1}) and (\ref{eq: SoundVelocity2}) and
solving the coupled equations, we obtain the first and second sound
velocities of a unitary Fermi gas, as reported in Fig. \ref{fig1_soundvelocity},
together with the LP ratio in the inset. In the previous work, the
two speeds of sounds were already calculated \cite{Arahata2009,Hu2010},
based on the GPF equations of state \cite{Hu2006} and the GPF superfluid
density \cite{Fukushima2007,Taylor2008}. The current update in Fig.
\ref{fig1_soundvelocity} improve the results near the transition
temperature $T_{c}\simeq0.167T_{F}$, because of the use of more accurate
equations of state. In particular, close to $T_{c}$ the improved
LP ratio is significantly smaller than the old result (see Fig. 3
in Ref. \cite{Hu2010}) and exhibits an interesting peak structure
at $T\sim0.15T_{F}\simeq0.9T_{c}$. Due to the smallness of the LP
ratio $\epsilon_{\textrm{LP}}$, the sound velocities are well approximated
by the expansion in terms of $\epsilon_{\textrm{LP}}$ \cite{Hu2010},
\begin{eqnarray}
c_{1}^{2} & = & v_{s}^{2}\left(1+\epsilon_{\textrm{LP}}x+\cdots\right),\label{eq:c12Approximate}\\
c_{2}^{2} & = & \frac{v^{2}}{\gamma}\left(1-\epsilon_{\textrm{LP}}x+\cdots\right),\label{eq: c22Approximate}
\end{eqnarray}
where 
\begin{equation}
x\equiv\frac{v^{2}}{\gamma v_{s}^{2}}.
\end{equation}
As shown in Fig. \ref{fig1_soundvelocity}, the approximate sound
velocities $c_{1}=v_{s}$ and $c_{2}=v/\sqrt{\gamma}$ are almost
identical to the exact results.

\begin{figure}[t]
\centering{}\includegraphics[width=0.48\textwidth]{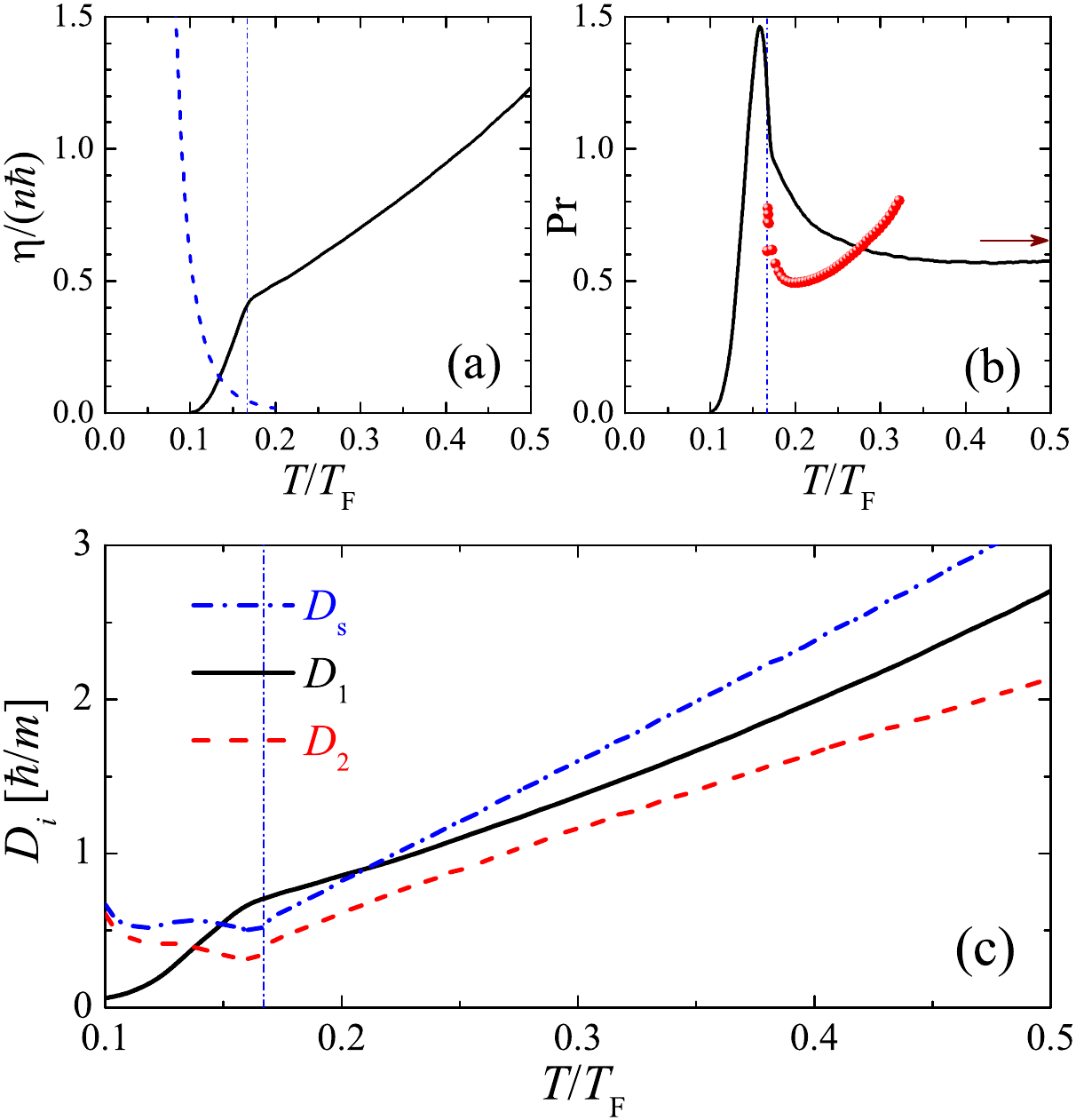}
\caption{\label{fig2_parameters} (color online). (a) Temperature dependence
of the shear viscosity of a unitary Fermi gas measured at NCSU \cite{Joseph2015}.
Above the superfluid transition, we use the refined experimental data
by Bluhm, Hou, and Schäfer \cite{Bluhm2017}. The blue dashed line
shows the shear viscosity in the deep superfluid phase, contributed
by the dominant four-phonon process (see Eq. (\ref{eq: ShearViscosityPhonons})).
(b) Temperature dependence of the Prandtl number calculated by using
the high-temperature approximate expression for thermal conductivity
(see Eq. (\ref{eq: ThermalConductivityHighTemperature})). The red
circles are the result of helium-II, with temperature rescaled according
to $(T/T_{c}^{\textrm{He}})\times0.167T_{F}$, where $T_{c}^{\textrm{He}}\sim2.1768$
K is the transition temperature of helium-II. The arrow to the right
shows the expected Prandtl number in the high-temperature limit, $\textrm{Pr}=2/3$.
(c) The various diffusion coefficients as a function of the reduce
temperature $\vartheta=T/T_{F}$. }
\end{figure}

To calculate the diffusion coefficients $D_{1}$, $D_{2}$ and $D_{s}$,
we note that the quantity $(\partial P/\partial T)_{\rho}$ can be
obtained by using the identity,
\begin{equation}
\left(\frac{\partial P}{\partial\rho}\right)_{s}=\left(\frac{\partial P}{\partial\rho}\right)_{T}+\left(\frac{\partial P}{\partial T}\right)_{\rho}\left(\frac{\partial T}{\partial\rho}\right)_{s}.
\end{equation}
For a unitary Fermi gas, a fixed entropy per unit mass $s$ means
$T/\rho^{2/3}$ is a constant and hence 
\begin{equation}
\left(\frac{\partial T}{\partial\rho}\right)_{s}=\frac{2}{3}\frac{T}{\rho}.
\end{equation}
Thus, it is easy to find that,
\begin{equation}
\frac{v^{2}}{v_{s}^{2}}\frac{2}{\rho s}\left(\frac{\partial P}{\partial T}\right)_{\rho}=\frac{3v^{2}\left(v_{s}^{2}-v_{T}^{2}\right)}{v_{s}^{2}Ts}=\frac{n_{s}}{n_{n}}\frac{3\epsilon_{\textrm{LP}}}{\gamma}\frac{f_{s}}{\vartheta f'_{s}}.
\end{equation}
By substituting this expression into the right hand side of Eq. (\ref{eq: DC2})
and defining the two dimensionless variables,
\begin{eqnarray}
A & = & f_{\eta}\left[\frac{4}{3}\frac{1}{n_{n}}+\frac{\gamma}{\textrm{Pr}}\right],\label{eq: AA}\\
B & = & f_{\eta}\left\{ \frac{4}{3}\left[x\gamma+\frac{n_{s}}{n_{n}}\left(1-\frac{3\epsilon_{\textrm{LP}}}{\gamma}\frac{f_{s}}{\vartheta f'_{s}}\right)\right]+\frac{1}{\textrm{Pr}}\right\} ,\label{eq: BB}
\end{eqnarray}
we then obtain,
\begin{eqnarray}
D_{1} & = & \frac{\hbar}{m}\left(\frac{Ac_{1}^{2}-Bv_{s}^{2}}{c_{1}^{2}-c_{2}^{2}}\right),\\
D_{2} & = & \frac{\hbar}{m}\left(\frac{Bv_{s}^{2}-Ac_{2}^{2}}{c_{1}^{2}-c_{2}^{2}}\right).
\end{eqnarray}
The remaining diffusion coefficient $D_{s}$ is given by, 
\begin{equation}
D_{s}=\frac{\hbar}{m}f_{\eta}\left[\frac{4}{3}\frac{n_{s}}{n_{n}}+\frac{\gamma}{\textrm{Pr}}\right].
\end{equation}

In Figs. \ref{fig2_parameters}a and \ref{fig2_parameters}b, we show
the universal function $f_{\eta}(\vartheta)$ and the Prandtl number
$\textrm{Pr}(\vartheta)$, respectively, by using the NCSU shear viscosity
data \cite{Joseph2015,Bluhm2017} and the high-temperature approximation
for thermal conductivity \cite{Braby2010}. The resulting diffusion
coefficients are reported in Fig. \ref{fig2_parameters}c. Near the
superfluid transition, all these coefficients are about $\hbar/m$,
indicating the strongly interacting nature of a unitary Fermi gas.

\subsection{The general structure of hydrodynamic dynamic structure factor}

Before we present the results of hydrodynamic dynamic structure factor,
it is useful to briefly discuss its general behavior, which is well
known in the literature \cite{Arahata2009,Hu2010,Hohenberg1965,Kadanoff2000,Mazenko2006}.
In the superfluid phase, the LP ratio $\epsilon_{\textrm{LP}}$ is
small, we may expand $D_{1}$ and $D_{2}$ in powers of $\epsilon_{\textrm{LP}}$
\cite{Hohenberg1965}:
\begin{eqnarray}
D_{1} & = & \frac{4\eta}{3\rho}+O\left(\epsilon_{\textrm{LP}}\right),\label{eq: D1SuperfluidPhase}\\
D_{2} & = & \frac{4\eta}{3\rho}\frac{\rho_{s}}{\rho_{n}}+\frac{\kappa}{\rho c_{p}}+O\left(\epsilon_{\textrm{LP}}\right).\label{eq: D2SuperfluidPhase}
\end{eqnarray}
At the leading order of $\epsilon_{\textrm{LP}}$, we obtain,
\begin{equation}
\frac{\chi_{nn}}{nk^{2}/m}\simeq\frac{Z_{1}}{\omega^{2}-c_{1}^{2}k^{2}+i\Gamma_{1}\omega}+\frac{Z_{2}}{\omega^{2}-c_{2}^{2}k^{2}+i\Gamma_{2}\omega},
\end{equation}
where $Z_{1}\equiv(c_{1}^{2}-v^{2})/(c_{1}^{2}-c_{2}^{2})$ and $Z_{2}\equiv(v^{2}-c_{2}^{2})/(c_{1}^{2}-c_{2}^{2})=1-Z_{1}$
\cite{Hu2010}, and $\Gamma_{1}\equiv D_{1}k^{2}$ and $\Gamma_{2}\equiv D_{2}k^{2}$.
The corresponding hydrodynamic dynamic structure factor then takes
the form,\begin{widetext}
\begin{align}
S\left(k,\omega\right) & =\frac{k^{2}}{2m}\frac{1}{\left(1-e^{-\beta\hbar\omega}\right)}\left[\frac{Z_{1}}{\pi}\frac{\Gamma_{1}\omega}{\left(\omega^{2}-c_{1}^{2}k^{2}\right)^{2}+\left(\Gamma_{1}\omega\right)^{2}}+\frac{Z_{2}}{\pi}\frac{\Gamma_{2}\omega}{\left(\omega^{2}-c_{2}^{2}k^{2}\right)^{2}+\left(\Gamma_{2}\omega\right)^{2}}\right]+O\left(\epsilon_{\textrm{LP}}\right),\\
 & \simeq\frac{k^{2}}{2m}\frac{1}{\left(1-e^{-\beta\hbar\omega}\right)}\frac{1}{\omega}\left[\frac{Z_{1}}{\pi}\frac{\Gamma_{1}/2}{\left(\omega-c_{1}k\right)^{2}+\left(\Gamma_{1}/2\right)^{2}}+\frac{Z_{2}}{\pi}\frac{\Gamma_{2}/2}{\left(\omega-c_{2}k\right)^{2}+\left(\Gamma_{2}/2\right)^{2}}+(\omega\rightarrow-\omega)\right],\label{eq: SkwSuperfluidState}
\end{align}
\end{widetext}featuring two sound waves at $c_{i}k$ with a damping
width $\Gamma_{i}=D_{i}k^{2}$ ($i=1,2$). It is easy to check that
the above dynamic structure factor satisfies both the compressibility
sum rule and the $f$-sum rule \cite{Nozieres1990,Griffin1993}, i.e.,
\begin{eqnarray}
\int_{-\infty}^{\infty}\frac{S(k,\omega)}{\omega}d\omega & = & \frac{1}{2mv_{T}^{2}},\\
\hbar^{2}\int_{-\infty}^{\infty}\omega S(k,\omega)d\omega & = & \frac{\hbar^{2}k^{2}}{2m}.
\end{eqnarray}
It is also straightforward to calculate the static structure factor
$S(k)=\hbar\int S(k,\omega)d\omega$. In the low-energy limit, where
$\hbar\omega\ll k_{B}T$, the static structure factor is approximated
by,
\begin{equation}
S\left(k\right)\simeq\frac{k_{B}T}{2m}\left[\frac{Z_{1}}{c_{1}^{2}}+\frac{Z_{2}}{c_{2}^{2}}\right]\equiv S_{1}\left(k\right)+S_{2}\left(k\right).
\end{equation}
Thus, the relative weight of second and first sound in $S(k,\omega)$
is given by \cite{Hu2010},
\begin{eqnarray}
\frac{S_{2}\left(k\right)}{S_{1}\left(k\right)} & = & \frac{Z_{2}/c_{2}^{2}}{Z_{1}/c_{1}^{2}}=\frac{v^{2}-c_{2}^{2}}{c_{1}^{2}-v^{2}}\frac{c_{1}^{2}}{c_{2}^{2}}\simeq\frac{\epsilon_{\textrm{LP}}}{1-v^{2}/v_{s}^{2}}.
\end{eqnarray}

As the temperature increases across the superfluid transition, it
seems that the second sound ceases to exist, as $Z_{2}$ becomes zero.
This is not fully correct, as in Eq. (\ref{eq: SkwSuperfluidState})
we neglect the terms at the order of $O(\epsilon_{\textrm{LP}})$.
What actually happens is that the propagating second sound mode turns
into a diffusive relaxation mode. To see this, we note that in the
normal state \cite{Hohenberg1965}, 
\begin{eqnarray}
D_{1} & = & \frac{4\eta}{3\rho}+\frac{\kappa}{\rho c_{p}}\epsilon_{\textrm{LP}},\label{eq: D1NormalState}\\
D_{2} & = & \frac{\kappa}{\rho c_{p}},\label{eq: D2NormalState}\\
D_{s} & = & \frac{\kappa}{\rho c_{v}}=D_{2}+D_{2}\epsilon_{\textrm{LP}},
\end{eqnarray}
and 
\begin{eqnarray}
\chi_{nn} & = & \frac{nk^{2}/m}{\omega^{2}-v_{s}^{2}k^{2}+i\Gamma_{1}\omega}+\chi_{nn}^{(2)},\\
\chi_{nn}^{(2)} & = & \frac{nk^{2}}{m}\frac{1}{\omega^{2}-v_{s}^{2}k^{2}+i\Gamma_{1}\omega}\frac{i\left(D_{s}-D_{2}\right)k^{2}}{\omega+i\Gamma_{2}}.
\end{eqnarray}
It is clear that $\chi_{nn}^{(2)}(k,\omega)$ peaks at $\omega=0$.
Thus, we may approximate 
\begin{equation}
\chi_{nn}^{(2)}\simeq-\frac{n}{mv_{s}^{2}}\frac{i\left(D_{s}-D_{2}\right)k^{2}}{\omega+i\Gamma_{2}}=-\frac{n}{mv_{s}^{2}}\frac{i\Gamma_{2}\epsilon_{\textrm{LP}}}{\omega+i\Gamma_{2}}.
\end{equation}
The corresponding dynamic structure factor $S^{(2)}(k,\omega)=-1/[\pi n(1-e^{-\beta\hbar\omega})]\textrm{Im}\chi_{nn}^{(2)}$
takes the form ($\omega\sim0$),
\begin{eqnarray}
S^{(2)}\left(k,\omega\right) & = & \epsilon_{\textrm{LP}}\left(\frac{k_{B}T}{mv_{s}^{2}}\right)\frac{\Gamma_{2}/\pi}{\omega^{2}+\Gamma_{2}^{2}},
\end{eqnarray}
which describes a \emph{thermally} diffusive mode of width $2\Gamma_{2}=2D_{2}k^{2}=2[\kappa/(\rho c_{p})]k^{2}$
\cite{Kadanoff2000,Mazenko2006}. The factor of $2$ in the width
comes from the fact that the second sound doublet below $T_{c}$,
each of width $\Gamma_{2}$, merges into a single central peak at
$\omega=0$.

\section{Collisionless vs hydrodynamic}

In this section, we present an estimation of the viscous relaxation
time (Fig. \ref{fig3_tinvtau}) and sketch out a sort of ``phase
diagram'', which at a given temperature determines the boundary between
the collisionless and hydrodynamic regimes in the plane of the transferred
momentum $k$ and the energy $\omega$ (Fig. \ref{fig4_phasediagram}). 

We note that, for a \emph{trapped} unitary Fermi gas, collective density
oscillations such as the breathing mode have been thoroughly studied
in the literature \cite{Kinast2004,Bartenstein2004,Hu2004,Altmeyer2007,Riedl2008,Tey2013},
by using the hydrodynamic theory. For these oscillations, the effective
momentum and energy are often much smaller than Fermi momentum and
Fermi energy, and hence the hydrodynamic description is applicable
without question \cite{HydrodynamicsInTrap}. In our case, the characteristic
momentum and energy of interest are about $0.1\sim1k_{F}$ and $0.1\sim1E_{F}$,
respectively. The condition of using the two-fluid hydrodynamic theory
then should be carefully examined.

\subsection{Viscous relaxation time}

In many cases, the boundary between the collisionless and hydrodynamic
domains might be determined by finding the average lifetime $\tau$
of the elementary quasi-particles that make the dominant contribution
to the thermodynamic and transport properties of the system considered.
For a unitary Fermi gas, this is difficult as the picture of well-established
quasi-particles may break down due to the inherent strong correlations.
In this work, we are lured into considering that the viscous relaxation
time related to the shear viscosity may be regarded as a characteristic
relaxation time scale. 

This consideration is inspired by the recent studies on the viscosity
spectral function $\eta(\omega)$, which is found to exhibit a clear
Drude peak of width $\hbar/\tau_{\eta}$ at zero frequency and a $\omega^{-1/2}$
tail at large $\omega$ \cite{Enss2011}, and satisfies an interesting
shear viscosity sum-rule at unitarity \cite{Taylor2010},
\begin{eqnarray}
\frac{2}{\pi}\int_{0}^{\infty}d\omega\left[\eta\left(\omega\right)-\frac{\hbar^{3/2}C}{15\pi\sqrt{m\omega}}\right] & = & P,
\end{eqnarray}
where $C$ is Tan's contact density and $P$ is the pressure. As a
result, it was shown that in the \emph{normal} state the viscosity
spectral function assumes the following form \cite{Enss2011},
\begin{equation}
\eta\left(\omega\right)\simeq\frac{P\tau_{\eta}}{1+\left(\omega\tau_{\eta}\right)^{2}}+\frac{\hbar^{3/2}C}{15\pi\sqrt{m\omega}}\frac{\omega\tau_{\eta}\left(1+\omega\tau_{\eta}\right)}{1+\left(\omega\tau_{\eta}\right)^{2}},\label{eq: ViscositySpectralFunctionDrudeForm}
\end{equation}
where the Drude weight in the first term on the right-hand-side of
the expression has to be the pressure $P$, in order to fulfill the
sum-rule. As $\omega\rightarrow0$, we thus obtain immediately the
useful viscosity-pressure relation,
\begin{equation}
\tau_{\eta}=\frac{\eta\left(\omega\rightarrow0\right)}{P}\equiv\frac{\eta}{P}.\label{eq: ViscosityPressureRelation}
\end{equation}
Indeed, this relation holds in the high-temperature limit, where quasi-particles
are well-defined and hence $\tau_{\eta}$ can be unambiguously calculated
using the kinetic theory \cite{Enss2012}. The argument presented
here indicates that the relation is applicable down to the superfluid
transition, near which quasi-particles may not be well-defined.

The viscosity-pressure relation can alternatively be understood from
the Einstein relation $\eta=\rho_{n}D_{\eta}$ that was first derived
by Hohenberg and Martin \cite{Hohenberg1965}. Here, $\rho_{n}$ is
the normal fluid density and $D_{\eta}\sim v_{eff}^{2}\tau_{\eta}$
is the diffusion coefficient related to shear viscosity. In the normal
state, $\rho_{n}=\rho=mn$ and the effective velocity $v_{eff}\sim v_{s}$.
By using the fact that $mv_{s}^{2}n\sim E/V\sim P$, we find the desired
relation $\eta\sim\tau_{\eta}P$.

\begin{figure}[t]
\centering{}\includegraphics[width=0.48\textwidth]{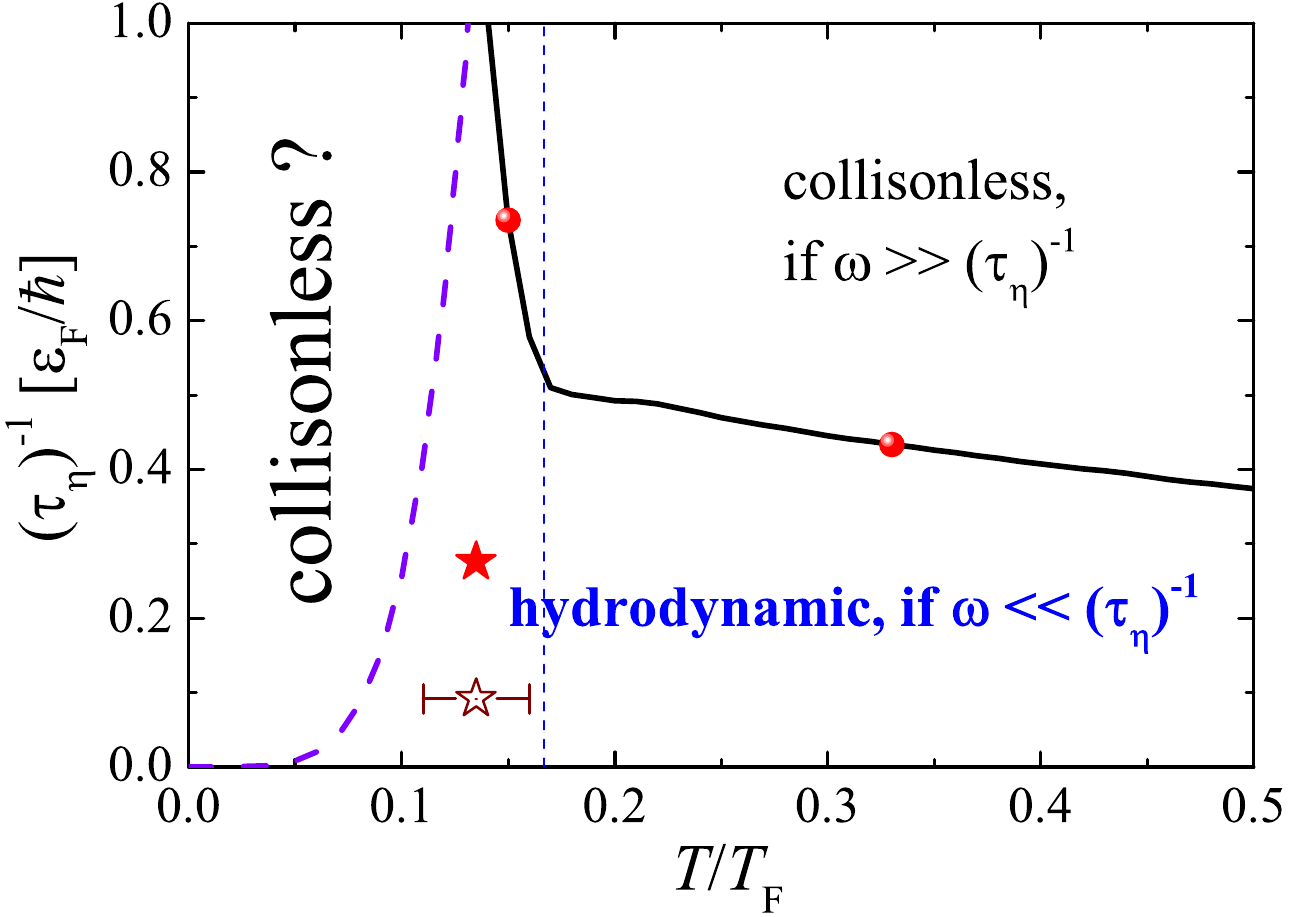}\caption{\label{fig3_tinvtau} (color online). The crossover frequency $\omega_{\eta}\equiv\tau_{\eta}^{-1}=P/\eta$
of a unitary Fermi gas as a function of temperature. The solid line
is obtained by using the NCSU data for shear viscosity \cite{Joseph2015,Bluhm2017}.
The purple dashed line is calculated using a low-temperature shear
viscosity contributed from the four phonon scattering process, by
\emph{assuming} phonons are well-defined quasi-particles at low temperature.
In both cases, for the pressure $P$ we take the MIT pressure equations
of state \cite{Ku2012}. The phase boundary at two selected temperatures
(red circles) is discussed in detail in Fig. \ref{fig4_phasediagram}.
The empty star indicates the frequency of second sound $\omega_{2}$
observed at Innsbruck \cite{Sidorenkov2013} and the solid star is
the correspondingly estimated low bound for $\omega_{\eta}\simeq3\omega_{2}\sim0.3\epsilon_{F}/\hbar$.}
\end{figure}

In the superfluid phase, the situation becomes complicated, since
the viscosity spectral function near zero frequency may deviate the
Drude form shown in Eq. (\ref{eq: ViscositySpectralFunctionDrudeForm}),
although the viscosity sum-rule should still be applicable. For simplicity,
however, we assume that this derivation is small and continue to use
the relation Eq. (\ref{eq: ViscosityPressureRelation}). It is also
worth noting that, experimentally it becomes difficult to measure
the shear viscosity in the low temperature regime (i.e., $T<0.1T_{F}$).
Theoretically, the only result of low-temperature shear viscosity
relies on the \emph{conjecture} that at that low temperature, phonons
make the dominant contribution to thermodynamic and transport properties.
The consideration of a four phonons scattering process leads to a
viscosity-entropy ratio \cite{Enss2011,Rupak2007}, 
\begin{equation}
\frac{\eta_{\textrm{ph}}}{s_{\textrm{ph}}}=\frac{\hbar}{k_{B}}2.15\times10^{-5}\xi^{5/2}\vartheta^{-8},
\end{equation}
where $\xi\simeq0.376$ is the Bertsch parameter \cite{Ku2012}. Using
the standard low-temperature expression $s_{\textrm{ph}}=(2\pi^{2}k_{B}/45)(k_{B}T/\hbar c_{1})^{3}$
and $c_{1}(T=0)=\sqrt{\xi/3}v_{F}$, we find that 
\begin{equation}
\frac{\eta_{\textrm{ph}}}{n\hbar}\simeq1.81\times10^{-4}\xi^{7/2}\vartheta^{-5},\label{eq: ShearViscosityPhonons}
\end{equation}
which is shown in Fig. \ref{fig2_parameters}a by a blue dashed line.

In Fig. \ref{fig3_tinvtau}, we report the crossover frequency $\omega_{\eta}\equiv\tau_{\eta}^{-1}=P/\eta$
as a function of temperature, calculated by using the MIT pressure
equation of state and the NCSU data for shear viscosity (black solid
line). Near zero temperature, where the NCSU data becomes very noisy,
we adopt the phonon expression Eq. (\ref{eq: ShearViscosityPhonons})
for the shear viscosity and plot the result with a purple dashed line.
The dynamics of the unitary Fermi gas is collisionless (hydrodynamic)
when $\omega\gg\omega_{\eta}$ ($\omega\ll\omega_{\eta}$). It is
not surprising that quite generally $\omega_{\eta}\sim O(\epsilon_{\eta}/\hbar)$,
since the unitary Fermi gas is strongly correlated and the Fermi energy
sets the characteristic energy scale close to the superfluid transition.
We observe a wide parameter window for the application of the two-fluid
hydrodynamic theory near the transition. In particular, at low temperature,
if the thermodynamics and dynamics of the unitary Fermi gas are dominated
by phonon excitations (which is to be confirmed experimentally yet),
we anticipate a crossover from the collisionless regime to the hydrodynamic
regime at $T\sim0.10T_{F}\simeq0.6T_{c}$, analogous to superfluid
$^{4}$He. For the latter, the crossover occurs at about $T\sim0.8\textrm{ K}\simeq0.4T_{c}^{\textrm{He}}$
\cite{Woods1973}. Of course, if the phonon assumption is not valid
at low temperature, the hydrodynamic region may extend down to zero
temperature.

To close this subsection, it is worth noting the second sound has
been observed at Innsbruck at $T=0.11-0.15T_{F}^{\textrm{trap}}$
under a sinusoidally modulation of repulsive laser beam at the trap
center \cite{Sidorenkov2013}. The modulation frequency $\omega_{2}$
is about $1720$ Hz or $0.092\epsilon_{F}/\hbar$, as shown in Fig.
\ref{fig3_tinvtau} by an empty star. The existence of the second
sound implies that $\omega_{2}\tau\ll1$. Thus, we must have $\tau^{-1}(T=0.135T_{F})\gg\omega_{2}\sim0.1\epsilon_{F}/\hbar$,
where we approximate $T_{F}^{\textrm{trap}}\sim T_{F}$. Naïvely,
we may estimate a \emph{low} bound for the crossover frequency 
\begin{equation}
\omega_{\eta}\left(T=0.135T_{F}\right)\simeq3\omega_{2}\sim0.3\frac{\epsilon_{F}}{\hbar}.
\end{equation}
This low bound is illustrated in Fig. \ref{fig3_tinvtau} by a solid
star. On the other hand, if we estimate the second sound velocity
$c_{2}(T=0.135T_{F})\sim0.08v_{F}$, we find that the wavevector of
the experimentally excited second sound is typically about $k=\omega_{2}/c_{2}\sim0.5k_{F}$. 

\subsection{Nature of sounds at two typical temperatures}

\begin{figure}[t]
\centering{}\includegraphics[width=0.48\textwidth]{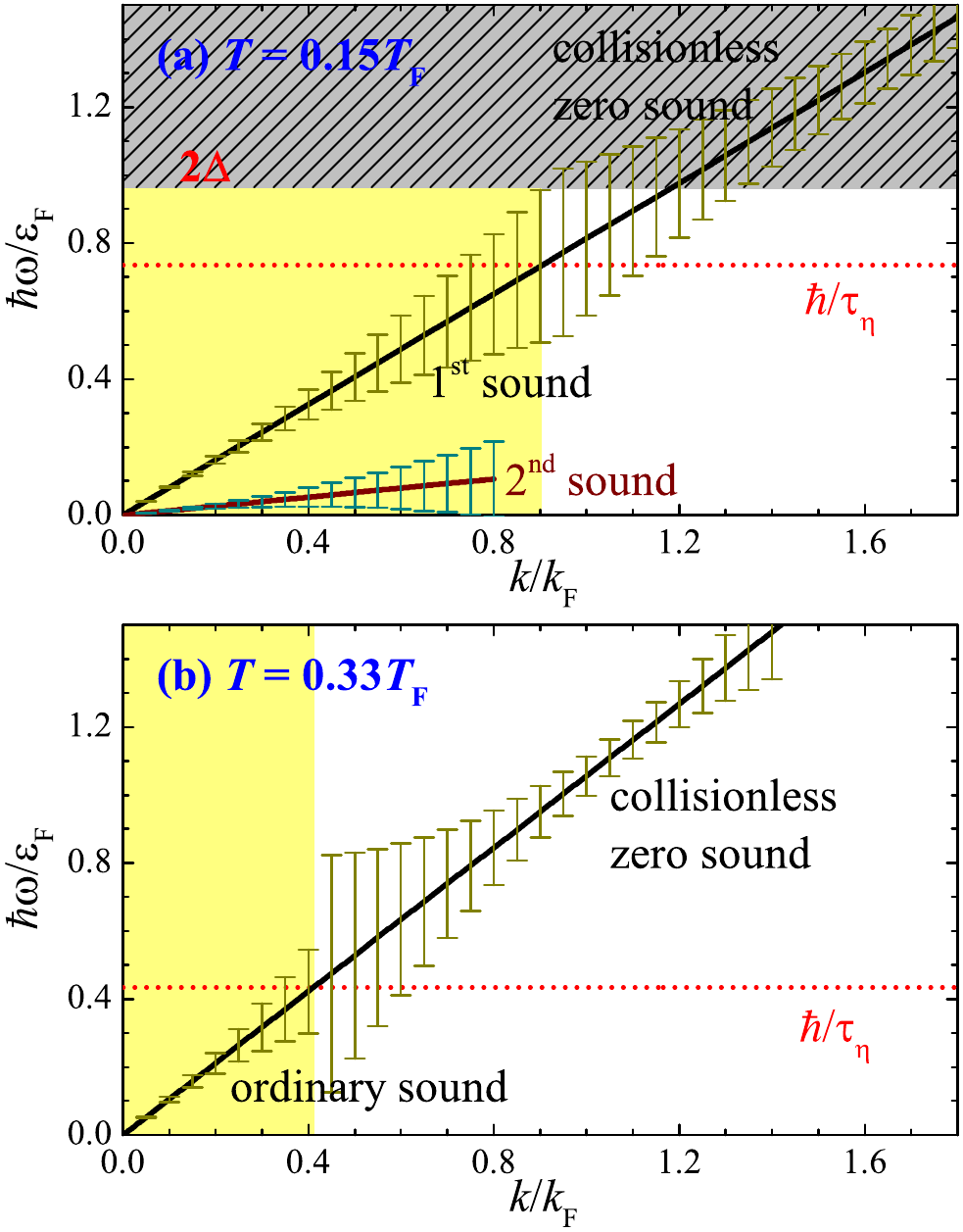}\caption{\label{fig4_phasediagram} (color online) Boundary between the hydrodynamic
and collisionless regimes below and above the superfluid transition
at two typical temperatures as highlighted in Fig. \ref{fig3_tinvtau}.
In the superfluid phase at $T=0.15T_{F}\simeq0.9T_{c}$ (a), the first
sound turns into a collisionless zero sound at $k\simeq0.9k_{F}$
and the second sound becomes a diffusive mode at $k\simeq0.8k_{F}$.
The shaded area shows the two-particle continuum at $\hbar\omega>2\Delta\sim\epsilon_{F}$.
In the normal phase at $T=0.33T_{F}\simeq2T_{c}$ (b), the hydrodynamic
ordinary sound turns into a collisionless zero sound at $k\sim0.4k_{F}$.
In both figures, the solid lines show $\omega_{i}=c_{i}(T)k$ ($i=1,2$)
and the error bars attached to the lines indicate the damping width
of the propagating sound modes.}
\end{figure}

Using the estimated crossover frequency $\omega_{\eta}=\tau_{\eta}^{-1}$,
we may qualitatively determine the collisionless-hydrodynamic boundary
as a function of the transferred momentum at a given temperature.
This is sketched in Fig. \ref{fig4_phasediagram}, for two typical
temperatures below and above the superfluid transition.

In the superfluid phase (a, $T=0.15T_{F}$), the crossover frequency
is large (i.e., $\omega_{\eta}\simeq0.7\epsilon_{F}/\hbar$), leading
to a significant characteristic momentum $k_{\eta}\simeq0.9k_{F}$.
Typically we find two propagating sound modes at low momentum $k\ll k_{\eta}$.
The hydrodynamic condition $\omega\ll\omega_{\eta}$ is always well
maintained for the second sound, due to its small frequency. However,
as the momentum increases, the rapidly increasing damping rate of
the second sound finally turns it into a diffusive mode, before reaching
$k_{\eta}$. For the first sound, it instead turns into a collisionless
zero sound once $k>k_{\eta}$. Here, we anticipate a large sound damping
due to the hydrodynamic to collisionless crossover, as sketched in
Fig. \ref{fig4_phasediagram}a at around $k\sim k_{\eta}$ or $\omega\sim\omega_{\eta}$.
By further increasing momentum, the collisionless zero sound enters
the two-particle continuum and is again damped via breaking Cooper
pairs or scattering off fermionic quasi-particles.

In the normal phase (b, $T=0.33T_{F}$), the crossover frequency is
also sizable (i.e, $\omega_{\eta}\simeq0.4\epsilon_{F}/\hbar$). At
this temperature, the second sound is already a thermal diffusive
mode, as we discuss earlier. The first sound or the ordinary sound
becomes the collisionless zero sound at $k\sim k_{\eta}\simeq0.4k_{F}$.

\section{Dynamic structure factor and sound attenuation at low momentum}

We are now ready to present the hydrodynamic dynamic structure factor
and understand the sound waves in the experimentally relevant parameter
space. In the following, we first consider the sound attenuation measurement
at MIT (where $k\apprle0.1k_{F}$) and then the Bragg scattering experiment
at SUT (where $k\sim0.5k_{F}$). At the end of this section, we finally
discuss the momentum dependence of the dynamic structure factor slightly
below the superfluid transition (with $T=0.15T_{F}$).

\begin{figure}[t]
\centering{}\includegraphics[width=0.48\textwidth]{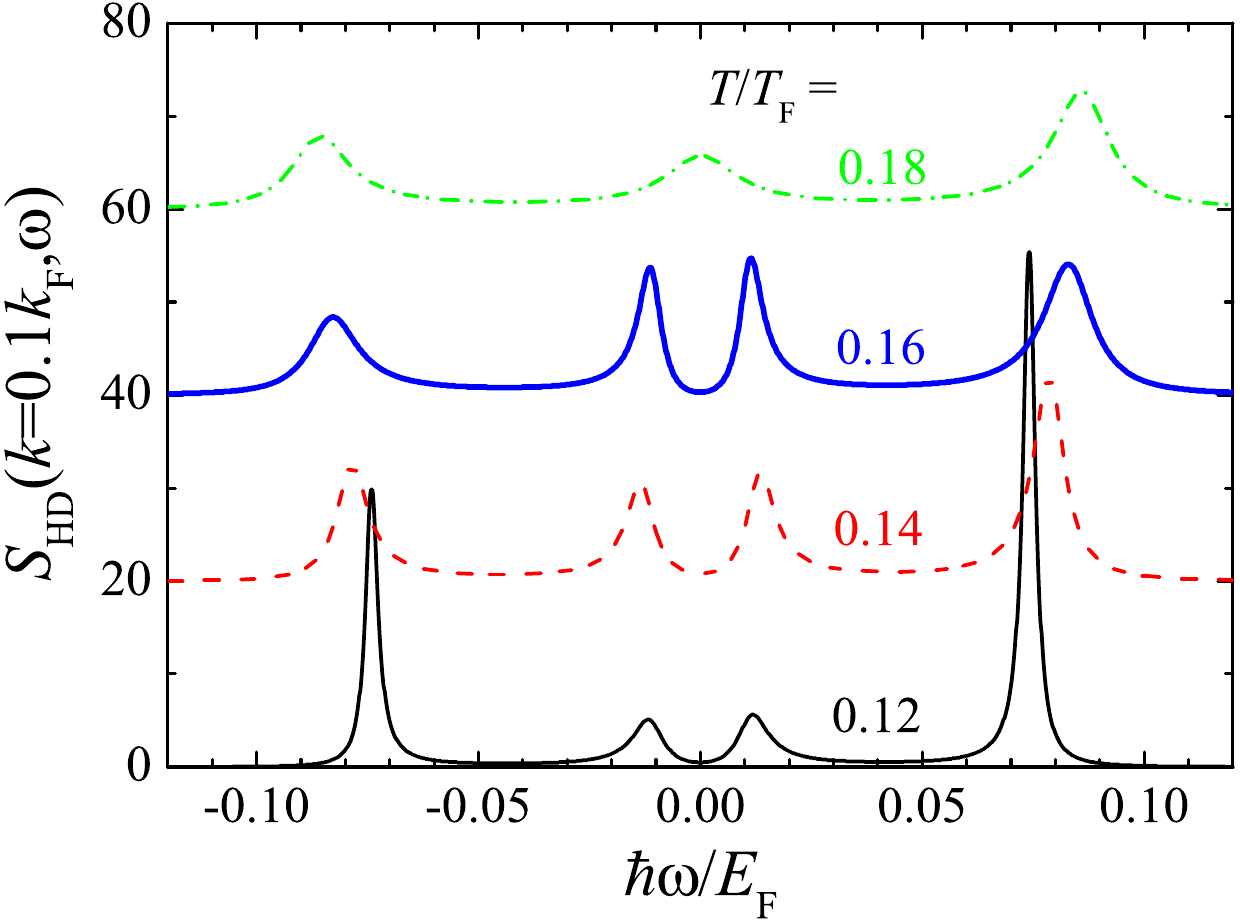}\caption{\label{fig5_dsf010kf} (color online) Temperature dependence of the
hydrodynamic dynamic structure factor at the transferred momentum
$k=0.1k_{F}$. The dynamic structure factor is measured in units of
$E_{F}^{-1}$.}
\end{figure}

\subsection{$k=0.1k_{F}$}

In Fig. \ref{fig5_dsf010kf}, we show the temperature evolution of
the dynamic structure factor at $k=0.1k_{F}$. At such a small transferred
momentum, we anticipate that the hydrodynamic condition may be well
satisfied for temperature $T\geq0.12T_{F}$. In the previous study,
the dynamic structure factor at the same transferred momentum was
calculated, using the non-dissipative two-fluid hydrodynamic theory
(see Fig. 4 in Ref. \cite{Hu2010}) and a temperature-independent
\emph{artificial} spectral broadening $\Gamma=0.002\epsilon_{F}/\hbar$.
Our updated results in Fig. \ref{fig5_dsf010kf} include the damping
effect and therefore remove the uncertainty in theoretical predictions.
The results can be directly compared with any experimental data once
they are available.

In the superfluid phase ($T<0.167T_{F}$), the two sound modes are
clearly visible. In particular, around zero frequency, we observe
a second sound doublet. With increasing temperature towards the superfluid
transition, the second sound becomes increasingly pronounced, as the
height of the sound peak increases. The width of the second sound,
however, is less dependent on temperature. As the width is roughly
given by $D_{2}k^{2}$, the temperature insensitivity of the width
may be understood from the fact that $D_{2}$ does not vary too much
with temperature, as can be seen from Fig. \ref{fig2_parameters}c.
Above the critical temperature, the second sound doublet merges into
a single broad Drude peak with width doubled (i.e., $2D_{2}k^{2}$),
as mentioned earlier.

\begin{figure}[t]
\centering{}\includegraphics[width=0.48\textwidth]{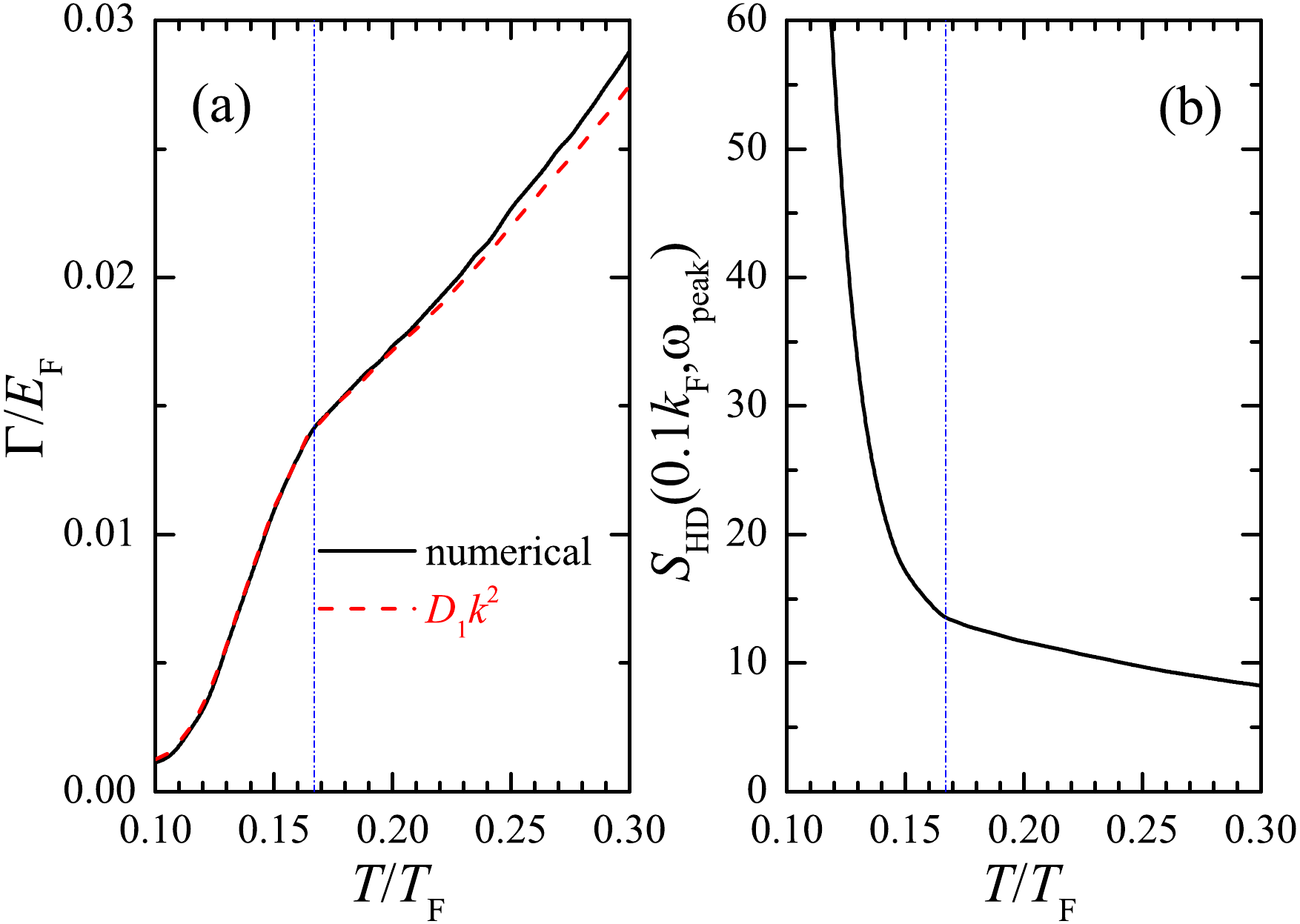}\caption{\label{fig6_1stdamping010kf} (color online) (a) Temperature dependence
of the damping width of the first sound at $k=0.1k_{F}$. The black
solid line is the calculated FWHM of the first sound mode. The red
dashed line is the anticipated damping rate $\Gamma=\Gamma_{1}=D_{1}k^{2}$.
(b) The peak value of the first sound (in units of $E_{F}^{-1}$)
as a function of temperature at $k=0.1k_{F}$. }
\end{figure}

The first sound, on the other hand, has a damping width that depends
critically on the temperature, as highlighted in Fig. \ref{fig6_1stdamping010kf}a.
The width follows closely the expression $D_{1}k^{2}$ as we anticipate.
Therefore, the temperature dependence of the width is a direct reflection
of the temperature dependence of the diffusion coefficient $D_{1}$,
which exhibits a sharp increase across the superfluid transition (see
again Fig. \ref{fig2_parameters}c). As a result, the peak height
of the first sound decreases rapidly across the critical temperature,
as shown in Fig. \ref{fig6_1stdamping010kf}b.

\subsection{$k=0.5k_{F}$}

\begin{figure}[t]
\centering{}\includegraphics[width=0.48\textwidth]{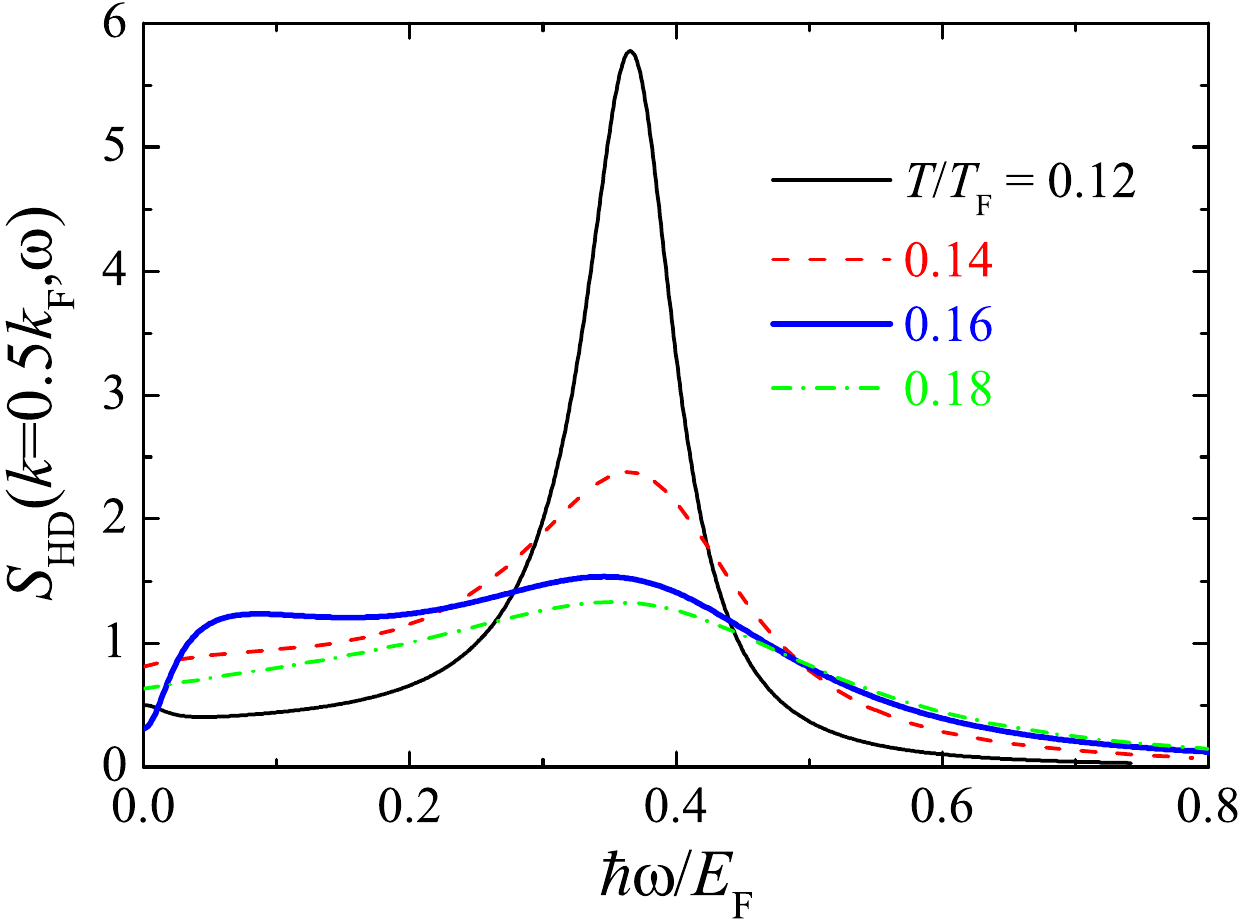}\caption{\label{fig7_dsf050kf} (color online) Temperature dependence of the
hydrodynamic dynamic structure factor at the transferred momentum
$k=0.5k_{F}$. We plot the dynamic structure factor at positive frequency
only. The dynamic structure factor at negative frequency can be obtained
by using the ``detailed balance'' relation, $S(k,-\omega)=e^{-\beta\hbar\omega}S(k,\omega)$
\cite{Nozieres1990,Griffin1993}. The dynamic structure factor is
measured in units of $E_{F}^{-1}$.}
\end{figure}

\begin{figure}[t]
\centering{}\includegraphics[width=0.48\textwidth]{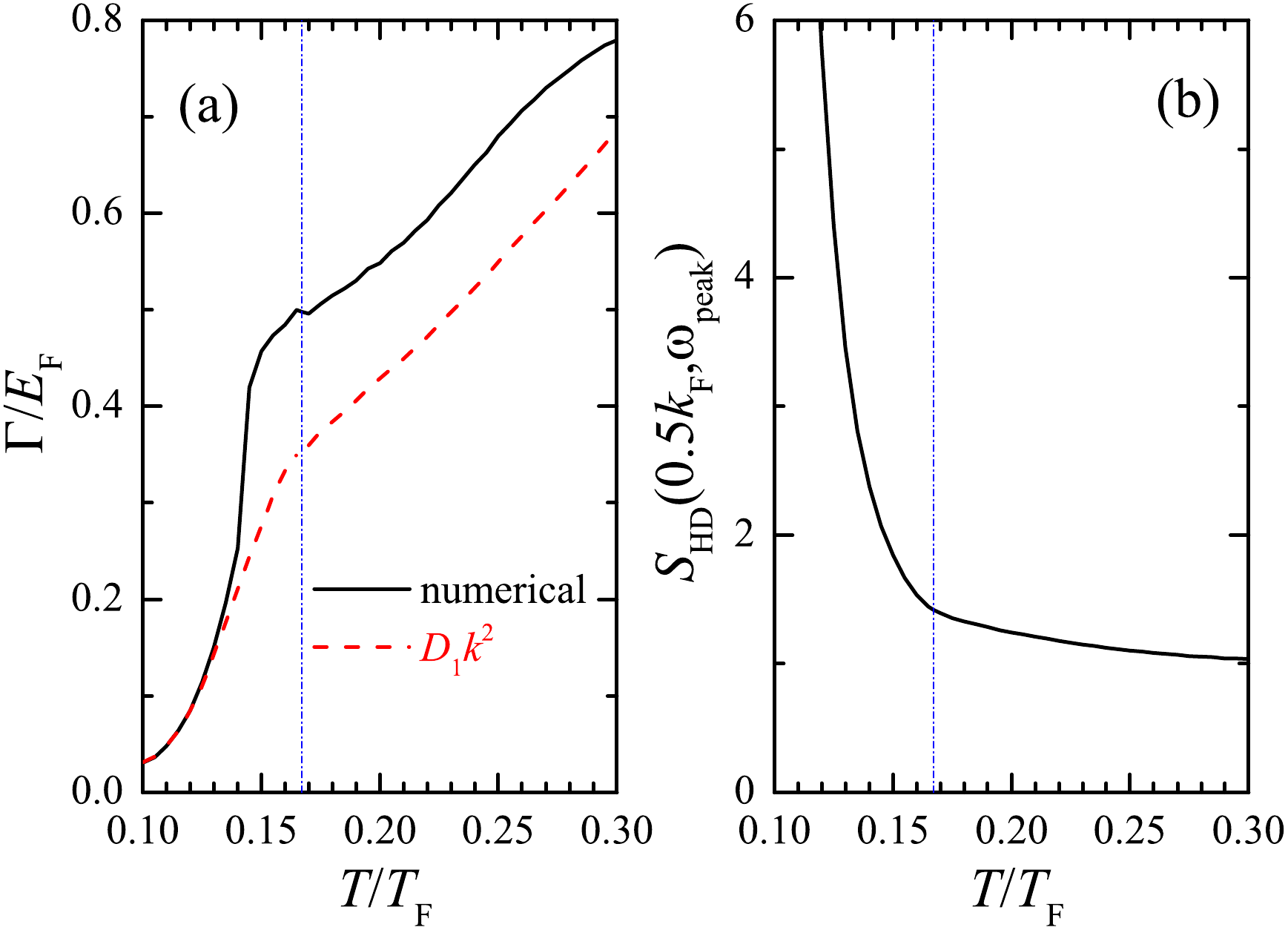}\caption{\label{fig8_1stdamping050kf} (color online) (a) The damping width
of the first sound as a function of temperature at $k=0.5k_{F}$.
The black solid line corresponds to the FWHM of the first sound mode
that is numerically determined. The red dashed line is the anticipated
damping rate $\Gamma=\Gamma_{1}=D_{1}k^{2}$, in the limit of small
damping $\Gamma\rightarrow0$. (b) The peak value of the first sound
(in units of $E_{F}^{-1}$) as a function of temperature at $k=0.5k_{F}$.}
\end{figure}

We consider now a moderately large transferred momentum $k=0.5k_{F}$,
which is of relevance to the SUT experiment. The hydrodynamic dynamic
structure factor is reported in Fig. \ref{fig7_dsf050kf} at several
temperatures across the superfluid transition. The corresponding damping
width and peak height of the first sound mode are shown in Figs. \ref{fig8_1stdamping050kf}a
and \ref{fig8_1stdamping050kf}b, respectively.

At this momentum, from Fig. \ref{fig4_phasediagram}a we find that
$\omega\tau_{\eta}\simeq0.6\sim1$ for the first sound and $\omega\tau_{\eta}\simeq0.1\ll1$
for the second sound at $T=0.15T_{F}$. Thus, the hydrodynamic condition
is marginally satisfied for the first sound and the relevant results
should be considered as qualitatively reliable only. On the other
hand, although the hydrodynamic condition is well fulfilled for the
second sound due to its small sound velocity, the damping rate of
$D_{2}k^{2}$ increases rapidly with $k$ and it may already be comparable
to the frequency $\omega=c_{2}k$ at $k\sim0.5k_{F}$. As a consequence,
the second sound may barely be seen in the dynamic structure factor,
although it \emph{was} observed from the sound wave propagation experiment
at similar wavevector as we have discussed at the end of Sec. IIIA.
Indeed, around the zero frequency we do not find interesting feature
at most temperatures, except at the temperature very close to the
transition temperature (i.e., at $T=0.16T_{F}$, the blue solid line),
where a very broad shoulder is observed. This broad shoulder should
be viewed as a remnant of the second sound.

We note that, the damping width for the first sound is also significant.
Actually, it is so significant at $T>0.14T_{F}$ that the width can
not be described by the expression $D_{1}k^{2}$ any more (see Fig.
\ref{fig8_1stdamping050kf}a), which is applicable only for small
damping rates. The peak height of the first sound decreases rapidly
across the superfluid transition (see Fig. \ref{fig8_1stdamping050kf}b),
similar to what happens in the case of $k=0.1k_{F}$.

\subsection{Momentum dependence of first and second sounds close to the transition}

\begin{figure}[t]
\centering{}\includegraphics[width=0.48\textwidth]{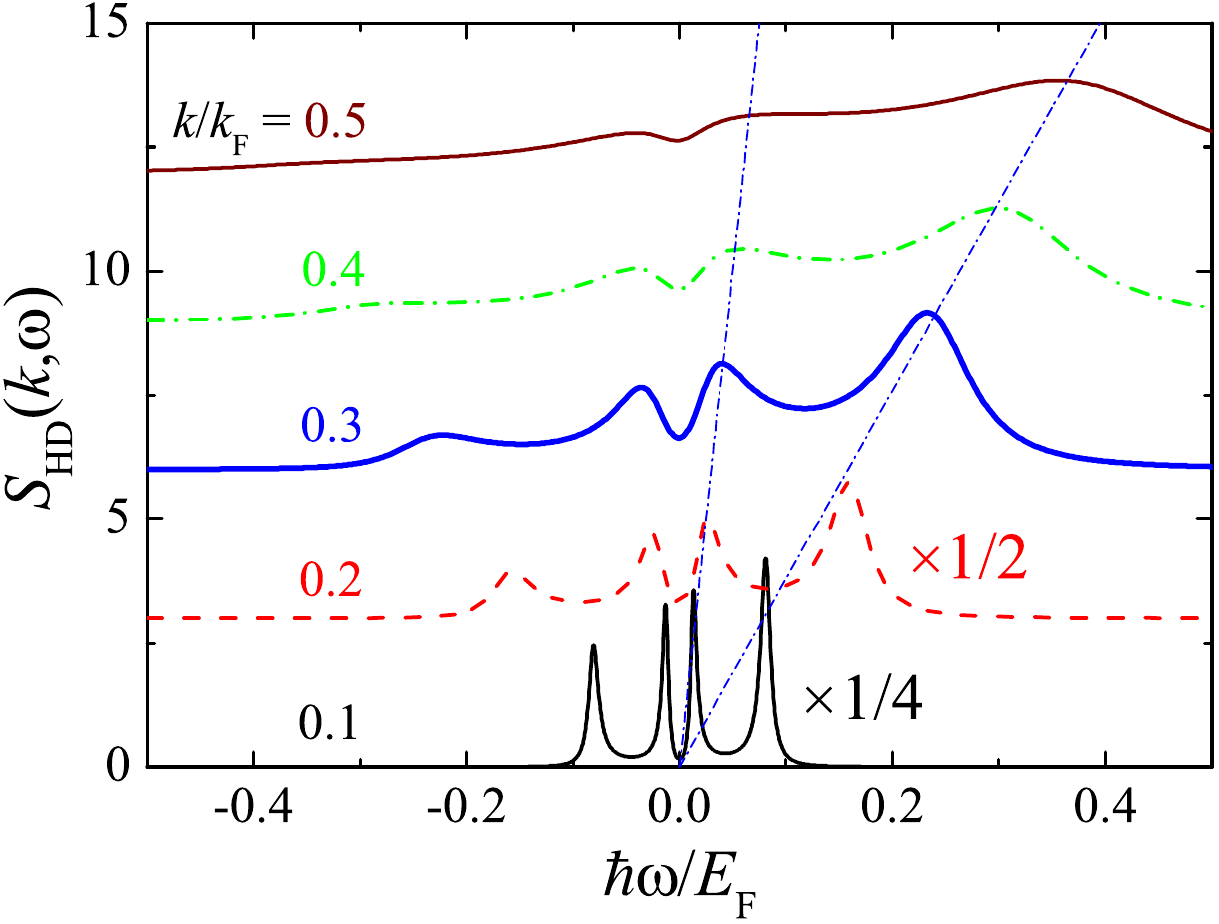}\caption{\label{fig9_dsfvaryk015tf} (color online) The hydrodynamic dynamic
structure factor as a function of the transferred momentum near the
superfluid phase transition ($T=0.15T_{F}$). The two dot-dashed lines
trace the peaks of the two sound modes. The value of the dynamic structure
factor at low momentum has been properly reduced for a better illustration.
The dynamic structure factor is measured in units of $E_{F}^{-1}$.}
\end{figure}

It is encouraging to find theoretically a remnant of the second sound
in dynamic structure factor slightly below the superfluid transition,
at the transferred momentum $k$ as large as $0.5k_{F}$. However,
we should bear in mind that, in Bragg scattering experiments, there
is an additional source for the spectral width of the sound modes,
the so-called \emph{instrumental} broadening, due to the finite duration
of the Bragg pulses. It is about $0.1\epsilon_{F}$ in the latest
Bragg scattering experiment \cite{Hoinka2017}. This additional broadening
may completely wash out the signal of the broad shoulder near $\omega=0$.
Therefore, experimentally it may be preferable to take a smaller transferred
momentum, although the small momentum may significantly reduce the
density response and hence make experimental data much more noisy.

In Fig. \ref{fig9_dsfvaryk015tf}, we show the hydrodynamic dynamic
structure factor as a function of the transferred momentum at $T=0.15T_{F}$.
It turns out that $k=0.3k_{F}$ could be an optimal choice for the
transferred momentum. On one hand, the damping width of the second
sound is reduced by a factor of $(5/3)^{2}\sim3$ and hence the second
sound can manifest itself clearly in the dynamic structure factor
(see the blue solid line). On the other hand, the hydrodynamic condition
is improved, as $\omega\tau_{\eta}\sim0.3$ becomes much smaller for
the first sound, compared with that in the case of $k=0.5k_{F}$.
Our two-fluid hydrodynamic description of both the first and second
sound may then be quantitatively reliable.

\section{Discussions}

We have now considered the dissipative two-fluid hydrodynamic theory
with a given set of superfluid density and transport coefficients
of a unitary Fermi gas. These inputs are collected in such a way that
they provide so far the state of the art that we can determine both
experimentally and theoretically. In this section, we would like to
discuss how will the results change, if we use different inputs for
superfluid density and thermal conductivity, both of which are less
understood in the literature. Here, the point is that, if the two
sound modes depend sensitively on superfluid density and thermal conductivity,
then we may determine them from the measured velocity and damping
width of sounds. 

\subsection{Dependence on the superfluid density}

\begin{figure}[t]
\begin{centering}
\includegraphics[width=0.48\textwidth]{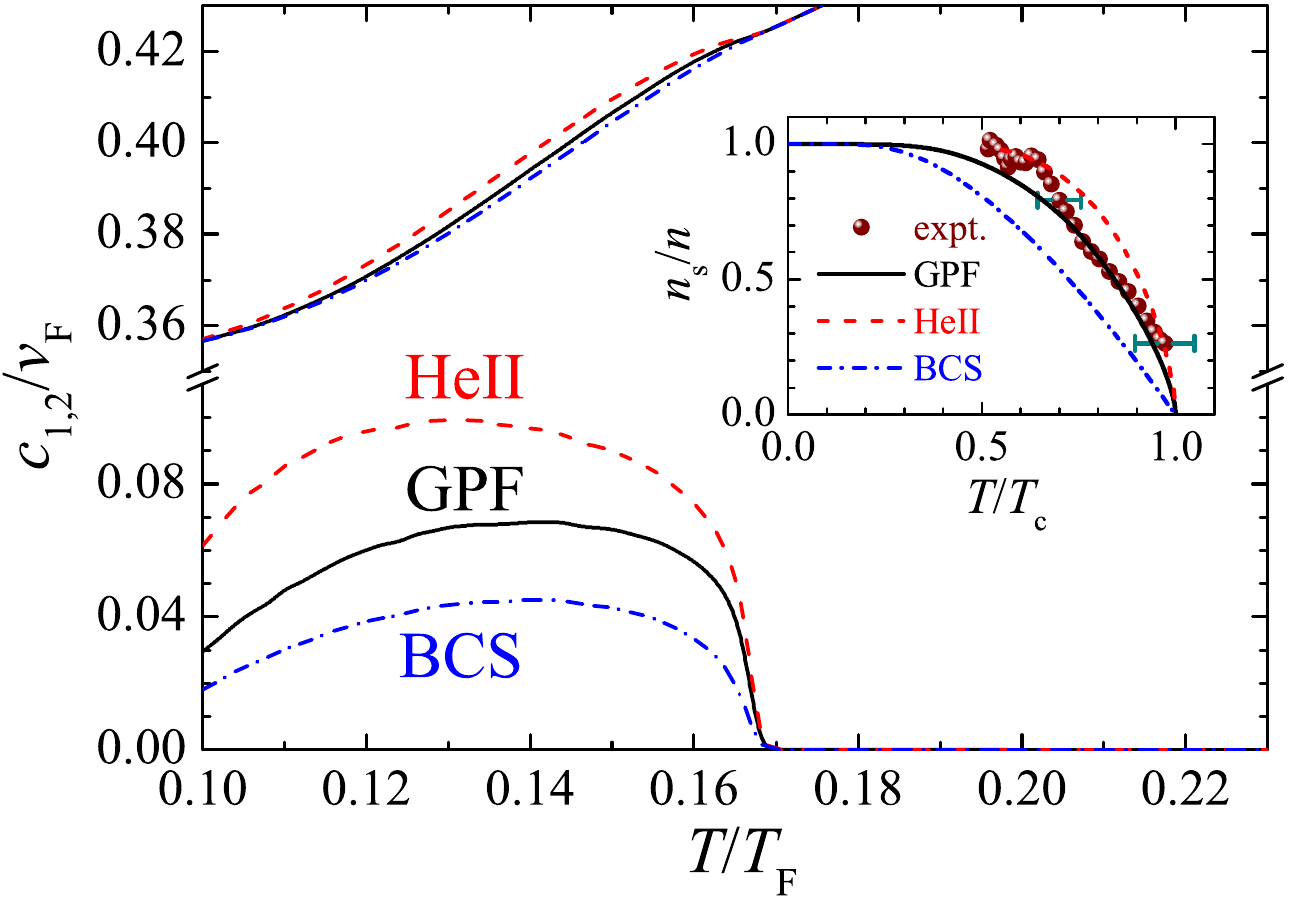}
\par\end{centering}
\centering{}\caption{\label{fig10_soundvelocityns} (color online) The first and second
sound velocities obtained with three different superfluid fractions
as plotted in the inset. In the inset, we show also the experimental
data of superfluid fraction (brown circles with error bars), recently
calibrated at Innsbruck \cite{Sidorenkov2013}.}
\end{figure}

A thorough discussion on the theoretical predictions of the superfluid
density in the unitary limit has been given in Ref. \cite{Taylor2008}.
Naïvely, we anticipate that the GPF theory provides so far the best
prediction. Indeed, it gives the best agreement with the measured
superfluid fraction at Innsbruck \cite{Sidorenkov2013}, as shown
in the inset of Fig. \ref{fig10_soundvelocityns}. For two possible
variants, we consider the superfluid fraction of superfluid $^{4}$He
\cite{Donnelly1998} and the superfluid fraction predicted by the
BCS mean-field theory. The resulting two sound velocities are shown
in the main figure of Fig. \ref{fig10_soundvelocityns}. 

It is apparent that the second sound velocity depends critically on
the superfluid fraction. This is easy to understand, since the square
of the second sound velocity can be accurately approximated by (see
Eq. (\ref{eq: c22Approximate}) and the red dashed line in the main
figure of Fig. \ref{fig1_soundvelocity}),
\begin{equation}
c_{2}^{2}\simeq T\frac{s^{2}}{c_{p}}\frac{\rho_{s}/\rho}{1-\rho_{s}/\rho}.
\end{equation}
As the entropy $s$ and the heat capacity $c_{p}$ have already been
accurately determined at MIT \cite{Ku2012}, the measurement of the
second sound velocity may provide a direct way to calibrate the superfluid
fraction of a unitary Fermi gas \cite{InnsbruckSuperfluidDensity}.

The first sound velocity, on the other hand, is only weakly affected
by the superfluid fraction via the coupling to the second sound. The
weak dependence is clear from the approximate sound velocity in Eq.
(\ref{eq:c12Approximate}), 
\begin{equation}
c_{1}^{2}\simeq v_{s}^{2}+\epsilon_{\textrm{LP}}c_{2}^{2}\simeq v_{s}^{2}+\epsilon_{\textrm{LP}}T\frac{s^{2}}{c_{p}}\frac{\rho_{s}/\rho}{1-\rho_{s}/\rho}.
\end{equation}

\begin{figure}
\begin{centering}
\includegraphics[width=0.48\textwidth]{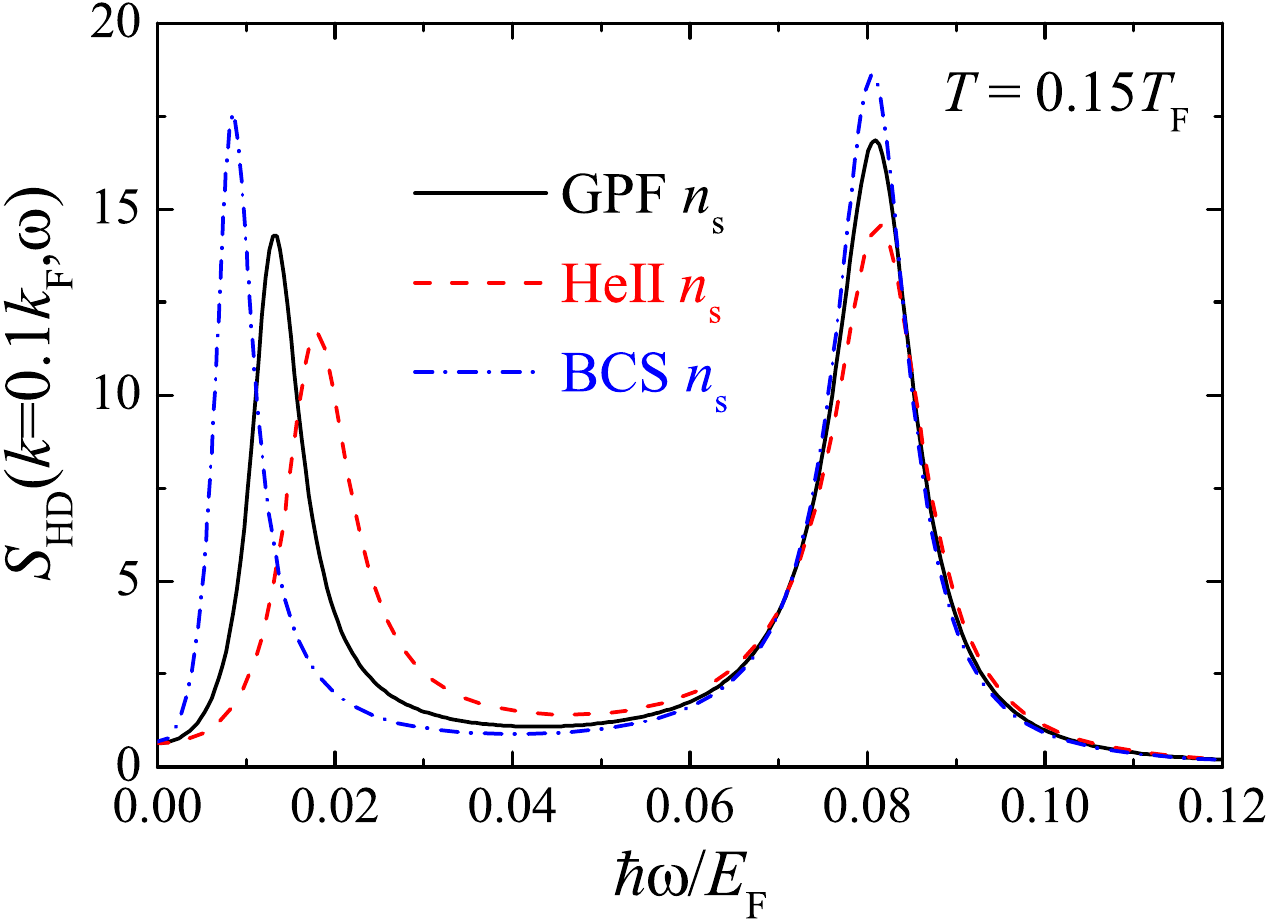}
\par\end{centering}
\centering{}\caption{\label{fig11_dsfns} (color online) The hydrodynamic dynamic structure
factor obtained with three different superfluid fractions. Here, we
take $k=0.1k_{F}$ and $T=0.15T_{F}$. The dynamic structure factor
is measured in units of $E_{F}^{-1}$.}
\end{figure}

In Fig. \ref{fig11_dsfns}, we present the dependence of the hydrodynamic
dynamic structure factor on the superfluid fraction, calculated with
$k=0.1k_{F}$ at $T=0.15T_{F}$. By changing the superfluid fraction,
the movement of the peak position of the two sound modes can be understood
from the change in the sound velocities. The damping width of the
two sounds becomes larger if we use a larger superfluid fraction (such
as that of superfluid $^{4}$He). In turn, it results in a smaller
peak height.

\subsection{Dependence on the thermal conductivity}

\begin{figure}
\begin{centering}
\includegraphics[width=0.48\textwidth]{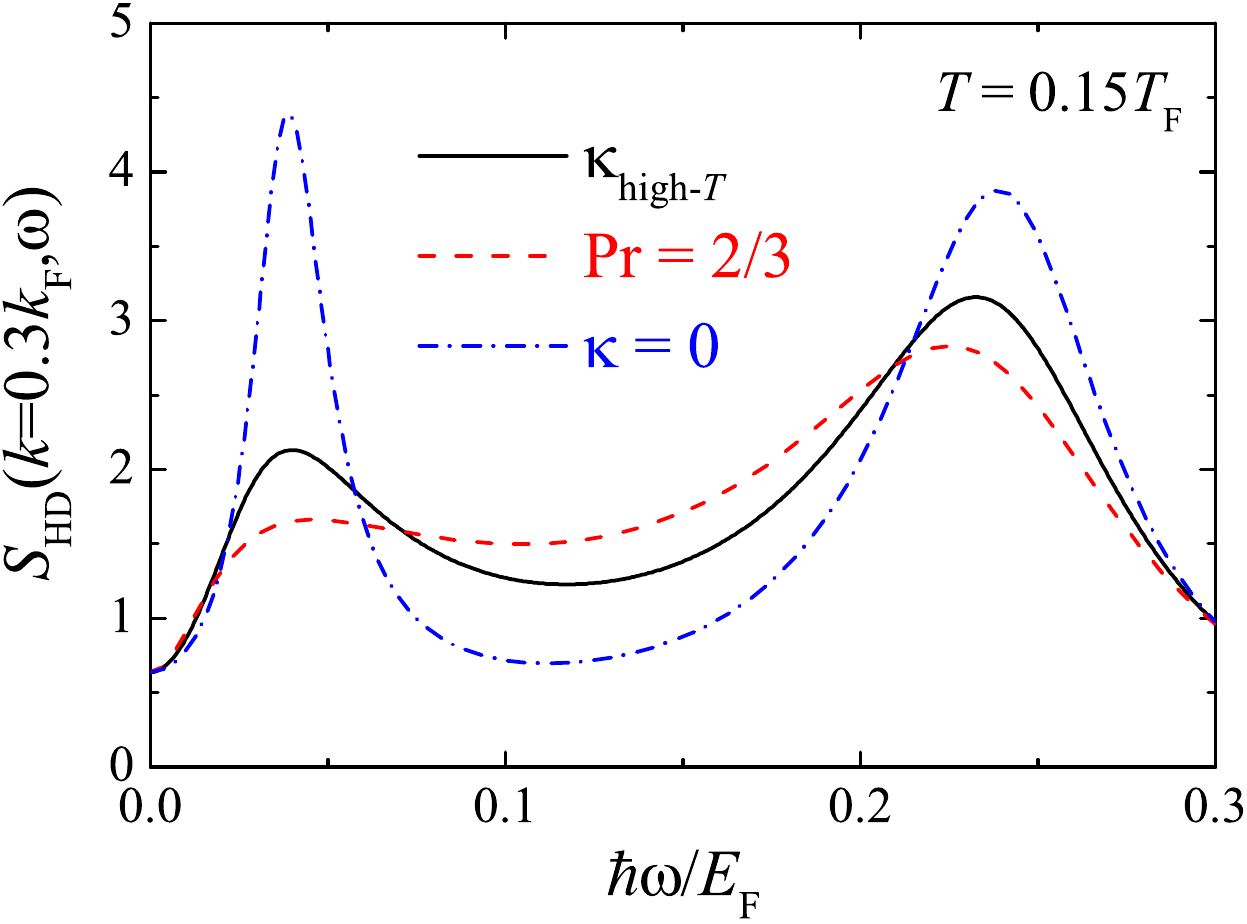}
\par\end{centering}
\centering{}\caption{\label{fig12_dsfthermalconductivity} (color online) The hydrodynamic
dynamic structure factor obtained with three different thermal conductivities:
the high-temperature approximation (black solid line), the thermal
conductivity that corresponds to a constant Prandtl number $\textrm{Pr}=2/3$
(red dashed line), and the zero thermal conductivity (blue dot-dashed
line). Here, we take $k=0.3k_{F}$ and $T=0.15T_{F}$. The dynamic
structure factor is measured in units of $E_{F}^{-1}$.}
\end{figure}

To understand the dependence of hydrodynamic dynamic structure factor
on the thermal conductivity, we consider a constant Prandtl number
$\textrm{Pr}=2/3$, in addition to the high-temperature approximation
for thermal conductivity that we have already used. The choice of
the factor of 2/3 is inspired by the fact that the strongly interacting
helium-II system has a Prandtl number $\textrm{Pr}\sim2/3$ at $T_{c}^{\textrm{He}}\leq T\leq2T_{c}^{\textrm{He}}$
\cite{Donnelly1998} (see, red circles in Fig. \ref{fig2_parameters}b).
We note also that in the high temperature limit the Prandtl number
is exactly $2/3$ \cite{Braby2010}. Another extreme limit that we
may choose is to simply set $\textrm{Pr}=\infty$. This is equivalent
to considering a vanishingly small thermal conductivity, $\kappa=0$.

The hydrodynamic dynamic structure factors with $k=0.3k_{F}$ and
$T=0.15T_{F}$ and at different choices of thermal conductivity are
reported in Fig. \ref{fig12_dsfthermalconductivity}. We do observe
a strong dependence of the dynamic structure factor on thermal conductivity.
In particular, if we do not take into account the effect of the thermal
conductivity ($\kappa=0$), the second sound peak becomes much narrower
and higher. This may be understood from the approximate expression
for the diffusion coefficient $D_{2}$ in the superfluid phase (see
Eq. (\ref{eq: D2SuperfluidPhase})). The first sound seems to be less
sensitive to the thermal conductivity than the second sound, since
the effect of thermal conductivity to the diffusion coefficient $D_{1}$
is weakened by a factor of the LP ratio $\epsilon_{\textrm{LP}}$,
according to Eq. (\ref{eq: D1SuperfluidPhase}).

\subsection{Transport coefficients from sound attenuation}

Ideally, if the damping width of the two sound modes, $W_{1}$ and
$W_{2}$, can experimentally measured at low momentum, we may directly
determine the shear viscosity and thermal conductivity of a unitary
Fermi gas, based on the known superfluid density and equations of
state. 

In the normal phase, the first sound width is $W_{1}=D_{1}k^{2}$.
The second sound is a thermally diffusive mode at $\omega=0$ and
its Drude width $W_{2}=2D_{2}k^{2}$. Using Eq. (\ref{eq: D1NormalState})
and Eq. (\ref{eq: D2NormalState}), we find that,
\begin{eqnarray}
\frac{\eta}{n\hbar} & = & \frac{3}{4}\frac{\left(W_{1}-\epsilon_{\textrm{LP}}W_{2}/2\right)}{\left(\hbar k^{2}/m\right)},\\
\frac{\kappa}{n\hbar} & = & \frac{1}{2}\frac{W_{2}}{\left(\hbar k^{2}/m\right)}c_{p}.
\end{eqnarray}

In the superfluid phase, the situation is a bit complicated, due to
the sound mode coupling. The damping widths of the first and second
sound are given by $W_{1}=D_{1}k^{2}$ and $W_{2}=D_{2}k^{2}$, to
the leading order of $O(\epsilon_{\textrm{LP}})$, respectively. From
the measured $W_{1}$ and $W_{2}$, we may calculate the two dimensionless
parameters $A$ and $B$ (see Eq. (\ref{eq: DC1}) and Eq. (\ref{eq: DC2})),
\begin{eqnarray}
A & = & \frac{W_{1}+W_{2}}{\left(\hbar k^{2}/m\right)},\\
B & = & \left(\frac{c_{2}}{v_{s}}\right)^{2}\frac{W_{1}}{\left(\hbar k^{2}/m\right)}+\left(\frac{c_{1}}{v_{s}}\right)^{2}\frac{W_{2}}{\left(\hbar k^{2}/m\right)}.
\end{eqnarray}
The shear viscosity $f_{\eta}\equiv\eta/(n\hbar)$ and the Prandtl
number $\textrm{Pr}\equiv\eta c_{p}/\kappa$ can then be obtained,
by solving the linearly coupled equations in Eq. (\ref{eq: AA}) and
(\ref{eq: BB}).

\section{Conclusions}

In summary, we have investigated the low-momentum dynamic structure
factor of a homogeneous unitary Fermi gas, from the viewpoint of a
dissipative two-fluid hydrodynamic theory. To this aim, we have estimated
the viscous relaxation time and have determined the characteristic
crossover frequency that distinguishes the collisionless region and
hydrodynamic region. Our estimation suggests that the dynamics of
the unitary Fermi gas is well described by the hydrodynamic theory
near the superfluid transition at the transferred momentum as low
as $0.5k_{F}$, where $k_{F}$ is the Fermi wavevector.

By collecting the best knowledge on the superfluid density, shear
viscosity and thermal conductivity, we have painted a general picture
of the hydrodynamic density response and have discussed in detail
the sensitive dependence of the two sound modes on the superfluid
density and thermal conductivity. The condition for observing the
second sound has been specifically addressed, in relation to the on-going
experiments at Swinburne University of Technology.

We have shown that the measurements of the velocity and damping width
of both first and second sound at sufficiently small momentum may
lead to an accurate determination of the superfluid density, shear
viscosity and thermal conductivity of a unitary Fermi gas. In this
respect, in the on-going experiment at Swinburne University of Technology,
if the Bragg scattering experiment can be carried out with $k\leq0.3k_{F}$
and the second sound can be successfully observed, then we will have
a very promising opportunity to improve the precision of the measured
superfluid density and shear viscosity and to determine the thermal
conductivity, which remains largely unknown both experimentally and
theoretically.
\begin{acknowledgments}
We thank very much Martin Zwierlein, Sascha Hoinka and Chris Vale
for useful discussions. This work was motivated by the sound attenuation
talk by Martin Zwierlein at the conference BEC 2017 held in Sant Feliu
de Guíxols, Spain and by Chris Vale's question of the conditions of
observing the second sound in his experimental setup (i.e., at a transferred
momentum $k\sim0.5k_{F}$). Our research was supported by Australian
Research Council's (ARC) Discovery Projects: FT130100815 and DP170104008
(HH), FT140100003 and DP180102018 (XJL), and by the National Natural
Science Foundation of China, Grant No. 11747059 (PZ).
\end{acknowledgments}

\end{document}